\documentclass[10pt, letter, twocolumn]{IEEEtran}
\usepackage{color}
\usepackage{times}
\usepackage{epsfig}
\usepackage{graphicx}
\usepackage{epstopdf}
\usepackage{algorithm}
\usepackage{algorithmic}
\usepackage{amsmath}
\usepackage{amssymb}
\usepackage{amsxtra}
\usepackage{multirow}
\usepackage{mathtools}
\usepackage{caption}
\usepackage{cleveref}
\usepackage{url}
\usepackage{subfigure}
\usepackage{cite}
\usepackage{subfigure}
\usepackage{amsthm}
\usepackage{setspace}
\newtheorem{theorem}{Theorem}
\newtheorem{lemma}{Lemma}

\newtheorem{corollary}{Corollary}

\newtheorem*{proof*}{Proof}

\newcommand{\RN}[1]{%
  \textup{\uppercase\expandafter{\romannumeral#1}}%
}

\begin{document}

\title{On the Secure and Reconfigurable Multi-Layer Network Design for Critical Information Dissemination in the Internet of Battlefield Things (IoBT)}

\author{ \IEEEauthorblockN{\large Muhammad Junaid Farooq, \textit{Student Member, IEEE} and Quanyan Zhu}, \textit{Member, IEEE}\\

\thanks{\vspace{0in}\hrule \vspace{0.2cm}
Preliminary results have been presented at the 15th Intl. Symposium on Modeling and Optimization in Mobile, Ad Hoc, and Wireless Networks (WiOpt 2017)~\cite{iobt_submitted}.

This research is partially supported by a DHS grant through Critical Infrastructure Resilience Institute (CIRI), grants CNS-1544782 and SES-1541164 from National Science of Foundation (NSF), and grant DE-NE0008571 from the Department of Energy (DOE). The statements made herein are solely the responsibility of the authors.

Muhammad Junaid Farooq and Quanyan Zhu are with the Department of Electrical \& Computer Engineering, Tandon School of Engineering, New York University, Brooklyn, NY, USA, E-mails: \{mjf514, qz494\}@nyu.edu.
}}

\maketitle

\begin{abstract}
The Internet of things (IoT) is revolutionizing the management and control of automated systems leading to a paradigm shift in areas such as smart homes, smart cities, health care, transportation, etc. The IoT technology is also envisioned to play an important role in improving the effectiveness of military operations in battlefields. The interconnection of combat equipment and other battlefield resources for coordinated automated decisions is referred to as the Internet of battlefield things (IoBT). IoBT networks are significantly different from traditional IoT networks due to battlefield specific challenges such as the absence of communication infrastructure, heterogeneity of devices, and susceptibility to cyber-physical attacks. The combat efficiency and coordinated decision-making in war scenarios depends highly on real-time data collection, which in turn relies on the connectivity of the network and information dissemination in the presence of adversaries. This work aims to build the theoretical foundations of designing secure and reconfigurable IoBT networks. Leveraging the theories of stochastic geometry and mathematical epidemiology, we develop an integrated framework to quantify the information dissemination among heterogeneous network devices. Consequently, a tractable optimization problem is formulated that can assist commanders in cost effectively planning the network and reconfiguring it according to the changing mission requirements. 
\end{abstract}

\IEEEpeerreviewmaketitle


\section{Introduction}
\begin{figure}
\centering
\includegraphics[width=3.0in]{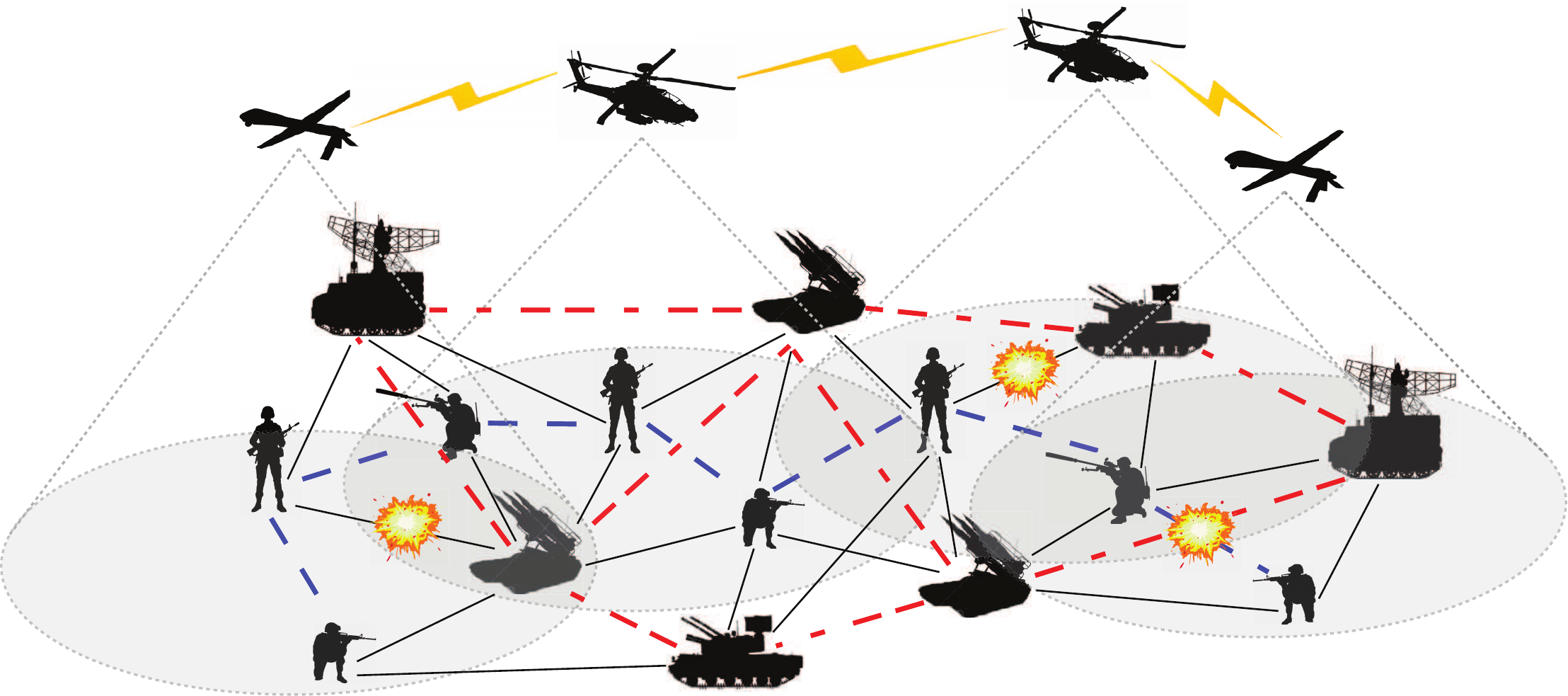}\\
\caption{A typical IoBT network with heterogeneous battlefield things and random enemy attacks. Battlefield devices interact with other devices within their communication range in a D2D manner for exchange of mission critical information. The link shape illustrates the piece of information being shared.
}\label{iobt_diagram}
\end{figure}

The Internet of things (IoT) is an emerging paradigm that allows the interconnection of devices which are equipped with electronic sensors and actuators~\cite{iot_ref1}. It allows for a higher level of situational awareness and effective automated decisions without human intervention. The concept has proven to be extremely useful in applications such as smart homes, energy management, smart cities, transportation, health care, and other domains~\cite{iot_ref2}. Recently, there is an interest in the defence community to leverage the benefits enabled by the IoT to improve the combat efficiency in battlefields and effectively manage war resources~\cite{iot_book,leveraging_iot_battle}. This emerging area of using IoT technology for defence applications is being referred to as the Internet of battlefield things (IoBT)~\cite{iobt_ref2,battle_things}. Fig.~\ref{iobt_diagram} illustrates a typical battlefield comprising of heterogeneous objects, such as soldiers, armoured vehicles, and aircrafts, that communicate with each other amidst cyber-physical attacks from the enemy.

The IoBT has the potential to completely revolutionize modern warfare by using data to improve combat efficiency as well as reduce damages and losses by automated actions while reducing the burden on human war-fighters. Currently, the command, control, communications, computers, intelligence, surveillance and reconnaissance (C$^4$ISR) systems use millions of sensors deployed on a range of platforms to provide situational awareness to military commanders and troops, on the ground, seas, and in the air~\cite{command_control}. However, the real power lies in the interconnection of devices and sharing of sensory information that will enable humans to make useful sense of the massive, complex, confusing, and potentially deceptive ocean of information~\cite{iot_review}. In the battlefield scenarios, the communications between strategic war assets such as aircrafts, warships, armoured vehicles, ground stations, and soldiers can lead to improved coordination, which can be enabled by the IoBT~\cite{iobt_ref1}. However, to become a reality, this vision will have to overcome several technical limitations of current information systems and networks.

Most civilian IoT applications such as smart homes and cities are infrastructure based, where the devices are connected to each other and the internet via an access point or gateway. In the battlefield scenario, the communication infrastructure such as cellular networks or base stations may not be available. Hence, the \emph{battlefield things} need to exploit device-to-device (D2D) communications~\cite{d2d_iot} to share information with other things\footnote{We use the terms ``things" and ``devices" interchangeably to refer to battlefield things throughout the paper.}. Therefore, the information sharing can be affected by the physical parameters of the network such as the transmission power of the things, the number of deployed things, their locations, and the flexibility of communication with other types of things. In addition to these factors, another impediment in the connectivity of battlefield things is the susceptibility to cyber-physical attacks. The information exchange between things may be affected by several factors such as jamming of radio frequency (RF) channels, physical attacks on infrastructure, node failures due to attacks on power sources, or lack of power, etc. Since the analytics and automated decisions in an IoBT network rely on the real-time data provided by the sensors deployed in the battlefield, we need to ensure the spread of information in the networks with a certain level of reliability and security to make accurate decisions.

Although the IoBT has to do to with much more beyond the connectivity of battlefield things, such as digital analytics and automated response, which allows the systems to react more quickly and precisely than humans; however, the connectivity aspect is vital in allowing the IoBT systems to unleash their full potential. It is ideal if we can achieve complete situational awareness and perfect information spreading throughout the network. However, due to limited available resources, incurred costs (capital and operational), and vulnerability to attacks, it is not practical and hence requires a cost-effective, secure and reconfigurable network design. Therefore, in this paper, we first characterize the information dissemination in an IoBT network under vulnerabilities, and use it to design the network parameters to reliably achieve mission specific performance goals with minimal amount of resources. We then present a reconfigurable framework that adapts with the changing connectivity situation of the network.

\subsection{Related Work}
Stochastic geometry (SG) based models have been successfully used in the modeling and analysis of traditional wireless networks such as cellular networks~\cite{andrews_coverage} and ad-hoc wireless sensor networks~\cite{haenggi_sg}. These models accurately capture the effect of spatial distribution of network devices on their connectivity and performance. However, they lack the capability of analysing the dynamics of information dissemination in communication networks.
One area of research related to this work is the study of probabilistic message broadcasting and percolation in random graphs~\cite{haenggi_sg,percolation}. Research in this direction has focused on identifying the conditions for the existence of giant connected components, and the analysis of connectivity.
On the other hand, mathematical epidemiology~\cite{epidemiology} has been studied extensively for analyzing the spread of viruses in computers, rumours in humans, and infectious diseases in biological networks~\cite{viruses,rumor,epidemic_scale_free}. Although the epidemic models are useful in capturing the diffusion of pathogens or rumors in a population, they do not take the geometry of the network into account and hence, cannot give meaningful insights in physically deployed communication networks. This research aims to investigate the dynamic decision-making and control of information dissemination and the design of multi-layer communication networks.
\\ \indent
Some attempts have been made to study the dynamical information dissemination in wireless networks such as~\cite{wireless_epidemics1,wireless_epidemics2}. However, they focus on traditional wireless adhoc networks, which do not capture the unique characteristics of IoBT networks. Moreover, there are few descriptive models available in literature for designing IoT networks, most of which are developed for civilian applications~\cite{hesham_iot}, and are not applicable to IoT networks over battlefields. Since IoBT networks are envisioned to exploit D2D communications and devices may be equipped with multiple radios for tactical communications, the information dissemination is directly associated with the communication characteristics of network devices as well as their associated deployment densities. Hence, it is imperative to develop an integrated design framework that can capture the specific features of IoBT networks.
\\ \indent
IoBT networks are composed of heterogeneous devices which may connect to more than one communication network based on the hardware capabilities. This requires a multi-layer network model that is able to incorporate the connectivity of such devices. In epidemiology literature, the multi-layer networks have been studied in the context of communication-contact layer networks~\cite{asymmetric2}. These are networks in which there is physical contact between the agents in one layer that allows for disease spreading and communication contact on the other layer, e.g., via telephone or social network, that allows for awareness spreading. The diffusion in such multiplex networks~\cite{multiplex1} is asymmetric and the interplay between the networks has been studied in~\cite{asymmetric1}. The dynamical process of multiple diseases spreading in multiplex networks is provided in~\cite{multiplex2}. However, our focus is on symmetric multi-layer communication networks that have different types of information propagating simultaneously in different network layers. Moreover, the physical connectivity in each layer has a certain structure based on the communication characteristics of the devices as opposed to complex networks that have non-trivial topological features. In this paper, we develop a specialized framework for information dissemination over IoBT networks that utilizes the connectivity structure in the multi-layer heterogeneous network by making use of the existing works in epidemiology. Some initial results of the developed framework have been presented in~\cite{iobt_submitted}.
However, the results are based on the assumption of reciprocity of communication between devices of one layer that are under the influence of the devices of other network layers. In this work, we lift this assumption and present a practical network model consisting of multi-radio devices that forms a multiplex communication network. We believe that such network models are well-suited to battlefield communications.

\subsection{Contributions}
In this paper, we develop a SG based model to characterize the connectivity of IoBT networks in terms of the degree distribution of the devices. We then use an epidemic spreading model to capture the dynamic diffusion of multiple messages within the network of devices at the equilibrium state. The resulting integrated open-loop system model is used as a basis for reconfiguring the network parameters to ensure a mission-driven information spreading profile in the network. This paper attempts to bridge the gap between the spatial stochastic models for wireless networks and the dynamic diffusion models in contact-based biological networks to derive new insights that aid in the planning and design of secure and reliable IoBT networks for mission critical information dissemination. The developed framework, with some modifications, is also applicable to the more general class of heterogeneous ad-hoc networks. The main contributions of this paper are summarized as follows:
\begin{itemize}
\item A novel multiplex network model for IoBT networks is proposed that helps in characterizing the intra-layer and network-wide connectivity of heterogeneous battlefield devices by considering the spatial randomness in their locations.
\item A tractable framework is developed for quantification of simultaneous information dissemination in the multiplex IoBT network based on mathematical epidemiology that considers the perceived level of threat to the network from cyber-physical attacks. Approximate closed form results relating the proportion of informed devices at equilibrium and the network parameters are provided.
\item An optimization problem is formulated that can assist military commanders in identifying the physical network parameters that are required in order to sufficiently secure the network from the perceived attacks. It can also help in reconfiguring existing networks to achieve a desired level of communication reliability.
\item A detailed investigation of the developed integrated framework is provided for particular battlefield missions and the effect of threat level and performance thresholds is studied.
\end{itemize}

%
%
%

\section{System Model}
In this section, we first describe the geometry of the IoBT network and propose a bi-layer abstraction model using tools from stochastic geometry. Then, we develop a dynamic model to characterize information dissemination in the heterogeneous IoBT network based on mathematical epidemiology.

\subsection{Network Geometry}
\begin{figure}
  \centering
  \includegraphics[width=3.3in]{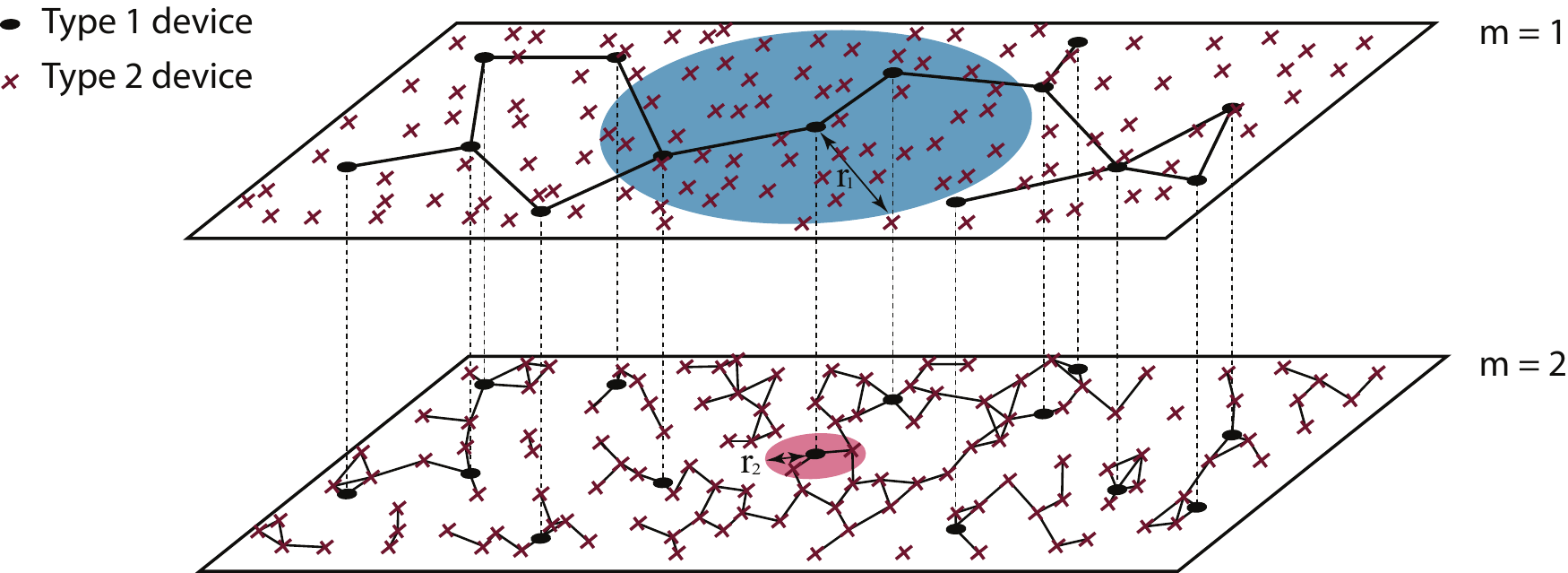}\\
  \caption{Heterogeneous IoBT network decomposed into virtual connectivity layers. The blue region illustrates the communication reach of type-$\RN{1}$ devices in layer 1 while the red region illustrates the communication reach of type-$\RN{1}$ and type-$\RN{2}$ devices in layer 2.}\label{sys_model}
\end{figure}

We consider uniformly deployed heterogeneous battlefield things in $\mathbb{R}^2$ that can be abstracted as a Poisson Point Process (PPP)\footnote{While the devices can be placed more strategically according to their characteristics and utility, however, due to potential mobility and difficulty of tracking network topology, these optimal locations may not be known or fixed, which justifies the PPP assumption.}~\cite{sg} with intensity $\lambda$ devices/km$^2$, referred to as $\Phi = \{\mathcal{X}_i, \mathcal{T}_i\}_{i \geq 1}$, where $\mathcal{X}_i$ and $\mathcal{T}_i$ represents the location and type of the $i^{th}$ device respectively. We assume that the network is composed of two types of devices, i.e., $\mathcal{T}_i \in \{1,2\}, \forall i \geq 1$. The first type of devices, i.e., $\mathcal{T}_i = 1$, referred to as \emph{type}-$\RN{1}$ devices, are equipped with two radio interfaces. The second type of devices, i.e., $\mathcal{T}_i = 2$, referred to as \emph{type}-$\RN{2}$ devices, have only one radio interface that is compatible with type-$\RN{1}$ devices. In the first network layer, only type-$\RN{1}$ devices can communicate with other type-$\RN{1}$ devices, while in the second layer, both type-$\RN{1}$ and type-$\RN{2}$ devices can communicate with each other due to the availability of common radios. Assuming that each device can be of type-$\RN{1}$ with probability $p$, i.e., $\mathbb{P}(\mathcal{T}_i = 1) = p, \forall i \geq 1$, then the set of active devices in the first network layer can be represented by a PPP $\Phi_1 = \{\mathcal{X}_i \in \Phi : \mathcal{T}_i = 1\}$ with intensity $\lambda_1 = p \lambda$, that is obtained by an independent thinning of the original point process $\Phi$. On the other hand, since all the devices in the second network layer can communicate with each other, so the active devices can be represented by a PPP $\Phi_2 = \Phi$, with intensity $\lambda_2 = \lambda$. \textcolor{black}{Note that the devices can move independently and we assume that the placement of devices as a result of mobility remains uncorrelated, i.e., can be represented by a new realization of the original PPP, thus resulting in a quasi-static network.}
\\ \indent
This type of network configuration is particularly well-suited to IoBT networks, where type-$\RN{1}$ devices may correspond to ground stations or armored vehicles equipped with multiple types of radios, while type-$\RN{2}$ devices may correspond to soldiers equipped with single radio smart mobile devices. The equipped radios on the devices are characterized by their transmission power or equivalently, the communication range $r_m \text{ in meters}$, $m \in \{1,2\}$. Type-$\RN{1}$ devices have two radios with transmission ranges $r_1$ and $r_2$, respectively, while type-$\RN{2}$ devices have one radio with transmission range $r_2$. The communication range of the radios is tunable in the interval $[r_m^{\min}$,$r_m^{\max}]$, where $r_m^{\min} \geq 0$ and $r_m^{\max} \geq r_m^{\min}, \forall m \in \{1,2\}$ \textcolor{black}{and we assume that $r_2 \leq r_1$.}
\\ \indent
Due to the absence of traditional communication infrastructure such as base stations, the devices are only able to communicate using D2D communications, i.e., devices $x,y \in \Phi_m$ are connected to each other in network layer $m$ if $\|x - y \| \leq r_m$, $m \in \{1,2\}$, where $\|.\|$ represents the Euclidean distance. Similarly, devices $x \in \Phi_1$ and $y \in \Phi \backslash \Phi_1$ can communicate with each other only if $\|x - y\| \leq r_2$.
Hence, the communication links between devices in each layer can be modeled using a random geometric graph (RGG)~\cite{rgg} with a connection radius of $r_m, m \in \{1,2\}$. Each of the layers of the multiplex network has a different connectivity that depends on the respective device densities and the communication ranges.
An illustrative representation of the network model is provided in Fig.~\ref{sys_model} that shows the connectivity between type-$\RN{1}$ devices in the first layer and the connectivity between all devices in the second layer.
In the subsequent subsection, we analyze the connectivity of the devices in each layer, referred to as \emph{intra-layer} connectivity and the connectivity of the overall network, referred to as \emph{network-wide} connectivity.

\subsection{Network Connectivity}\label{connectivity_sec}
In this subsection, we describe the connectivity between the heterogeneous devices in an IoBT network. The connectivity of devices can be classified into \emph{intra-layer} and \emph{network-wide} connectivity, which are explained as follows:
\subsubsection{Intra-layer Connectivity}
Within a particular network layer $m$, the active\footnote{Type-$\RN{1}$ devices are active in both network layers while type-$\RN{2}$ devices are only active in the second network layer.} devices can communicate with each other if they are within a distance of $r_m$ of each other. Referring to Fig.~\ref{iobt_diagram}, the dotted lines between battlefield vehicles represent the connectivity of the devices in the first layer. Similarly, the solid lines between soldiers and vehicles represent the connectivity of the devices in the second layer. The set of communication neighbours of a typical device $x \in \Phi_m$ in layer $m$, can be expressed as $\mathcal{N}_m(x) = \{y \in \Phi_m : \|x-y\| \leq r_m\}$. The connectivity of the RGG formed by devices in layer $m$ is characterized by the degree of the devices, denoted by $K_m$, which is defined as the number of neighbours of each device, i.e., $K_m = |\mathcal{N}_m(x)|$, where $|.|$ represents the set cardinality.
Since the network is spatially distributed as a PPP, the degree of each device in the RGG is a Poisson random variable~\cite{haenggi_sg}. Therefore, the resulting intra-layer degree distribution of a typical device can be expressed by the following lemma.
\begin{lemma} \label{lemma_intra_layer_degree}
The intra-layer degree of a typical device in each network layer is distributed as follows:
{
\begin{align} \label{K_1_pdf}
\mathbb{P}(K_1 = k) =
\left\{
	\begin{array}{ll}
		(1 - p) + pe^{-\lambda_1 \pi r_1^2}, & \mbox{if } k = 0, \\
		pe^{-\lambda_1 \pi r_1^2}\frac{(\lambda_1 \pi r_1^2)^k}{k!}, & \mbox{if } k > 0,
	\end{array}
\right.
\end{align}}
{
\begin{align}\label{K_2_pdf}
\mathbb{P}(K_2 = l) = e^{-\lambda_2 \pi r_2^2}\frac{(\lambda_2 \pi r_2^2)^l}{l!}, \ l \geq 0,
\end{align}}
for sufficiently large $\lambda_1, \lambda_2, r_1$ and $r_2$. The average degree of a typical device in the two network layers can be expressed as follows:
{
\begin{align}
\mathbb{E}[K_1] = p \lambda_1 \pi r_1^2, \label{E_K1}\\
\mathbb{E}[K_2] = \lambda_2 \pi r_2^2. \label{E_K2}
\end{align}}
\textbf{\emph{Proof.}} See \textbf{Appendix~\ref{proof_lemma_intra_layer_degree}}.
\end{lemma}
From Fig.~\ref{sys_model}, it is clear that the average degree, or equivalently the connectivity of devices in each layer depends on the density of the deployed devices as well as the communication range.
\textcolor{black}{The joint probability distribution of the connectivity of a typical device in each layer is denoted by $\mathbb{P}(K_1 = k, K_2 = l)$.}

\subsubsection{Network-wide Connectivity}

If the two network layers are collapsed together to form a single virtual network such that both layers reinforce the connectivity of the devices, then the connectivity is characterized in terms of the combined degree denoted by $K_c$. The combined-layer degree of a typical device $x \in \Phi$ is defined as $K_c = |\mathcal{N}_1(x)| + |\mathcal{N}_2(x)|$. Since the degree of the devices in each layer is Poisson distributed, the combined-layer degree follows a Poisson mixture distribution expressed by the following lemma.
\begin{lemma}\label{combined_degree_lemma}
The average combined-layer degree of a typical device can be expressed as follows:
{
\begin{align} \label{E_K_c}
\mathbb{E}[K_c] = p \lambda_1 \pi r_1^2 + \lambda_2 \pi r_2^2.
\end{align}}
\textbf{\emph{Proof.}} See \textbf{Appendix~\ref{combined_degree_lemma_proof}}
\end{lemma}
In the following section, we describe the information dissemination over such bi-layer networks that have the above mentioned connectivity profile.

\section{Information Dissemination} \label{sec:dynamics}
Each type of device in the IoBT network generates data that needs to be propagated to other devices of the same type and/or different types of devices depending on the role of that device. There are certain pieces of information that needs to be shared among the same type of devices, e.g., soldiers need to communicate information with other soldiers and similarly commanding units might also share information amongst them. Henceforth, we refer to the information sharing in each network layer as \emph{intra-layer information dissemination}.
On the other hand, some information might be important for all network nodes such as network health monitoring data or network discovery beacons. This is henceforth referred to as \emph{network-wide information dissemination}.
We assume a time slotted system where the duration of each slot is $\tau$ s. During each time slot, the informed devices broadcast information to their neighbors at a rate of $\gamma$.
Let $\mathcal{P}_s^{(i)} \in [0,1]$ be the average probability that the transmitted information type $i \in \{1,2,c\}$ by a typical device is successfully received by its neighbors, referred to as the \emph{success probability}, and $\delta \in [0,1]$ be the probability that the communication will be affected by cyber-physical attacks. Since the the event of successful transmission due to interference from other devices and the event of a cyber-physical attack are independent, therefore the effective probability of a successful transmission can be expressed as $(1 - \delta)\mathcal{P}_s^{(i)}$. Consequently, the information spreading rate between devices, denoted by $\alpha^{(i)}, i \in \{1,2,c\}$, can be expressed as follows:
\begin{align}
\alpha^{(i)} = \gamma (1 - \delta) \mathcal{P}_s^{(i)},
\end{align}
We refer to $\delta$ as the \emph{threat level} as it signifies the perceived risk in information transmission between devices. Without loss of generality\footnote{There is no loss of generality since $\tau$ can be made arbitrarily small.}, we can select the contact rate $\gamma = 1$, so effectively, $\alpha^{(i)}$ is the probability of successful information transmission between devices. In essence, $\alpha^{(i)}$ can be interpreted as the desired security level from the perceived threats to the communication network. \textcolor{black}{The probability of successful transmission can be computed by setting a threshold on the received signal-to-interference-plus-noise-ratio (SINR) at a typical device. Several techniques can be used from SG literature to accurately characterize the SINR based success probability depending on the medium access protocol used~\cite{SG_book_Bartek}. We use a generalized representation of the success probability $\mathcal{P}_s^{(i)} = g_i(p, \lambda, \mathbf{r})$ in terms of the densities and communication ranges of the devices, where the function $g_i(.)$ is assumed to be monotone in its arguments.} The parameter $\delta$ can capture a broad range of cyber-physical threats in IoBT networks. Different methods can be used to assess the threat level in battlefields due to jamming, physical attacks, and other adversarial actions based on historical data and/or statistical models of attack types, some of which are explored in existing works such as~\cite{jamming} and~\cite{attacks2}. For instance, to model jamming attacks, the parameter $\delta$ can be based on the SINR, in which case the RGG becomes an interference graph~\cite{stirg}. To tackle physical network attacks such as targeted attacks, the parameter $\delta$ can be based on the density of device deployment, the connectivity of devices, or the type of devices. Furthermore, an integrated metric can also be developed that can simultaneously capture a multitude of threats. However, developing such a comprehensive metric is beyond the scope of the current work and we use a generic threat level for illustrative purposes. Over time, the adversarial attacks may compromise or negatively impact significant portion of the network connectivity. So there is a need for a resilient framework that can be reconfigured to recover from the lost connectivity by cyber-physical attacks.

Another important aspect of the information dissemination process is to account for the information annihilation at each time step. There are several reasons that a device may not broadcast the information that it has received from another device in the previous time slot such as limited buffer capacity and misclassifying information as unimportant. However, the most important factor is to ensure propagating the most recent information in the network. This dynamical information spreading process in an IoBT network can be formalized using the susceptible-infected-susceptible (SIS) model~\cite{epidemics}, which is well studied in mathematical epidemiology. The challenge is that information propagates over a topology in wireless communication networks, while the classical SIS model does not deal with topological constraints. Furthermore, we deal with simultaneous information dissemination in multiple network layers which presents additional challenges.
In the following subsection, we describe the dynamics of the information dissemination process.


\subsection{Information Dynamics}
In this section, we present the dynamics of information dissemination across the IoBT network. For ease of explanation, we first describe the dynamics of a single message being propagated in the network, followed by the dynamics of two messages simultaneously spreading in the network.
\subsubsection{Single Message Propagation}
If all the devices in the network disseminate the same message from one device to another in a broadcast manner during each time slot using all the available radio interfaces, then each device with degree $k$ can either be in an uninformed state ($U_k$) or an informed state ($I_k$) depending on the success of information transmission. To model this behaviour and explain the dynamics of information dissemination across the IoBT network, we exploit the susceptible-infected-susceptible (SIS) model~\cite{epidemics} from mathematical epidemiology. The information dissemination is directly related to the degree of the devices in the network, which in turn depends on the physical network parameters. Since the network is random with potentially a large number of devices, we use the degree based mean-field approach, in which all devices are considered to be statistically equivalent in terms of the degree and the analysis is done on a typical device. Therefore, the information dissemination dynamics of the system can be written in terms of the degree of the devices as follows~\cite{contact_process}:
\begin{align} \label{differential_equation}
\frac{d I^{(c)}_{k}(t)}{dt} = - \mu I^{(c)}_{k}(t) + \alpha^{(c)} k U^{(c)}_{k}(t) \Theta^{(c)},
\end{align}
where $I^{(c)}_k(t)$ denotes the proportion of devices with degree $k$ that are in state $I_k$, i.e., informed with network-wide information, at time $t$, $U^{(c)}_{k}(t) = 1 - I^{(c)}_{k}(t)$ denotes the proportion of degree $k$ devices that are in state $U_k$. The superscript $(c)$ refers to the combined-layer network signifying network-wide information dissemination. The first term in \eqref{differential_equation} explains the annihilation of information with time, \textcolor{black}{i.e., the informed devices return to the uninformed sate at a rate of $\mu$. This ensures that at each time step, only the most recent information is propagated in the network as a particular piece of information is discarded after being retransmitted multiple times depending to the annihilation rate. For equilibrium analysis, we can assume that $\mu = 1$  as the effect of the annihilation can be captured by the effective spreading rate.}
The second term accounts for the creation of informed devices due to the spreading. The rate of increase in the density of informed devices with degree $k$ is directly proportional to the degree, the probability of successful transmission of information $\alpha$, the proportion of uninformed devices with degree $k$, i.e., $U_{k}^{(c)}(t)$, and the average probability that a neighbour of a device with degree $k$ is informed, denoted by $\Theta^{(c)}$. \textcolor{black}{The probability $\Theta^{(c)}$ can be computed as $\sum_{k^\prime} \mathbb{P}(K_c^{\text{neighbour}} = k^\prime | K_c = k) I^{(c)}_{k^\prime}(t)$, where $\mathbb{P}(k^\prime | K_c = k)$ denotes the probability that a neighbour of a typical device with $K_c = k$ has a degree $K_c^{\text{neighbour}} = k^\prime$. Since the network is PPP, the degrees are uncorrelated, i.e., $\mathbb{E}[K_c^{\text{neighbour}} K_c] = \mathbb{E}[K_c^{\text{neighbour}}] \mathbb{E}[K_c]$, so we can effectively write $\mathbb{P}(K_c^{\text{neighbour}} = k^\prime | K_c = k) I^{(c)}_{k^\prime}(t) = \frac{k^{\prime}\mathbb{P}(K_c^{\text{neighbour}} = k^\prime)}{\mathbb{E}[K_c]}I^{(c)}_{k^\prime}(t)$.
Hence, by appropriately renaming dummy variables, $\Theta^{(c)}$ can be expressed as follows~\cite{epidemics}:}
\begin{align} \label{theta_expression}
\Theta^{(c)} = \sum_{k \geq 0} \frac{k \mathbb{P}(K_c = k)}{\mathbb{E}[K_c]} I^{(c)}_{k}(t),
\end{align}
where $\mathbb{E}[K_c]$ is provided in Section~\ref{connectivity_sec}. Note that another expression for the rate of change in the uninformed devices, i.e.,$\frac{d U^{(c)}_{k}(t)}{dt}$, can also be written; however, it is useless since $U^{(c)}_{k}(t)$ depends directly on $I^{(c)}_{k}(t)$ and thus \eqref{differential_equation} completely describes the dynamics of network-wide information dissemination.
\begin{figure}
  \centering
  \includegraphics[width=2.4in]{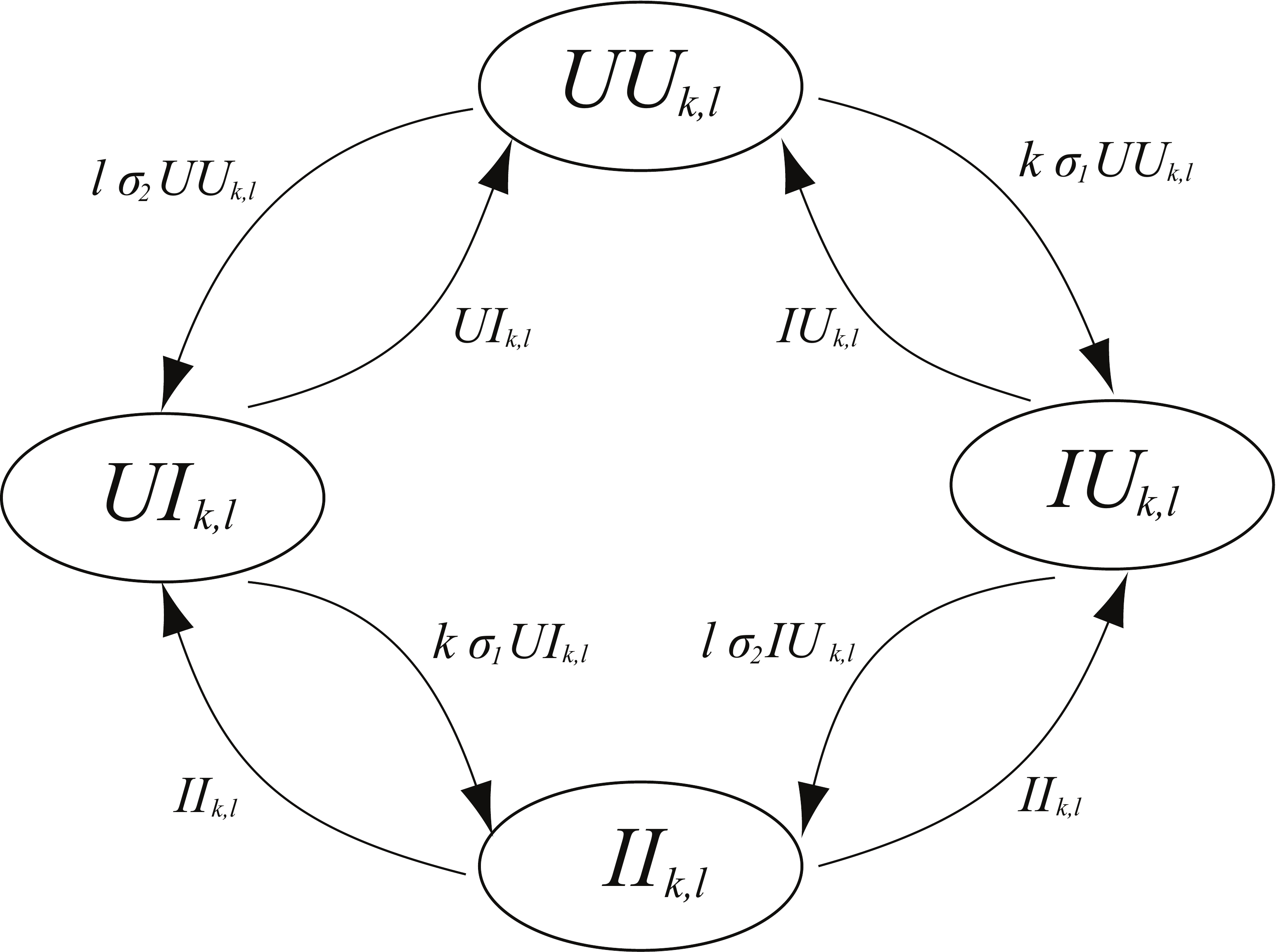}\\
  \caption{State transition diagram for the simultaneous diffusion of two different messages in the IoBT network. At each time instant, a change in status of only one of the messages is allowed. The arrows are labeled with the transition probabilities.}\label{transitions}
\end{figure}

\subsubsection{Multiple Message Propagation}
In general, there may be multiple messages or pieces of information spreading in the IoBT network at any particular time. However, for the bi-layer network model, we assume that there are two messages propagating in the network, i.e., one in each network layer. Therefore, each device with a degree $k$ in the first layer and $l$ in the second layer can be in one of the four possible dynamical states, i.e., uninformed of both messages ($UU_{k,l}$), informed of message 1 but uninformed of message 2 ($IU_{k,l}$), uninformed of message 1 but informed of message 2 ($UI_{k,l}$), and informed of both messages ($II_{k,l}$). These state variables denote the proportion of devices in the network that are in that particular state. To model the coupled dynamics of this information diffusion process, we can make use of the SIS-SIS interaction model~\cite{multiplex2} from mathematical epidemiology. A state transition diagram is given by Fig.~\ref{transitions}. Notice that a change in status of one of the messages is allowed at each time instant. The set of differential equations describing the state evolution are given by \eqref{diff1} - \eqref{diff4},
\begin{figure*}[t!]
\begin{align}
&\frac{dUU_{k,l}(t)}{dt} = - (\alpha^{(1)} k \Theta_1 + \alpha^{(2)} l \Theta_2) UU_{k,l}(t) + IU_{k,l}(t) + UI_{k,l}(t), \label{diff1} \\
&\frac{dIU_{k,l}(t)}{dt} = \alpha^{(1)} k \Theta_1 UU_{k,l}(t) - (\alpha^{(2)} l \Theta_2 + 1)IU_{k,l}(t) + II_{k,l}(t), \label{diff2} \\
&\frac{dUI_{k,l}(t)}{dt} = \alpha^{(2)} l \Theta_2 UU_{k,l}(t) - (\alpha^{(1)} k \Theta_1 + 1)UI_{k,l}(t) + II_{k,l}(t), \label{diff3} \\
&\frac{dII_{k,l}(t)}{dt} = \alpha^{(1)} k \Theta_1 UI_{k,l}(t) + \alpha^{(2)} l \Theta_2 IU_{k,l}(t) - 2II_{k,l}(t), \label{diff4}
\end{align}
\end{figure*}
where $\Theta_1$ and $\Theta_2$ are the effective probabilities of a completely uninformed device, i.e., in state $UU_{k,l}$, to get informed by message 1 and message 2 respectively. They can be evaluated as follows:
\begin{align}
&\Theta_1 = \frac{1}{\mathbb{E}[K_1]} \sum_{k,l} \mathbb{P}(K_1 = k,K_2 = l) k \left(IU_{k,l,\sigma_1,\sigma_2}(t) \ + \right. \notag \\ & \hspace{2.2in} \left. II_{k,l,\sigma_1,\sigma_2}(t) \right), \label{eq_sigma1}\\
&\Theta_2 = \frac{1}{\mathbb{E}[K_2]} \sum_{k,l} \mathbb{P}(K_1 = k,K_2 = l) l \left(UI_{k,l,\sigma_1,\sigma_2}(t) \ + \right. \notag \\ & \hspace{2.2in} \left. II_{k,l,\sigma_1,\sigma_2}(t) \right), \label{eq_sigma2}
\end{align}
Since $UU_{k,l}(t) + UI_{k,l}(t) + IU_{k,l}(t) + II_{k,l}(t) = 1$, so in essence, only three of the four differential equations are linearly independent. The particular quantity of interest is $II_{k,l}(t)$ as it signifies the proportion of devices that are informed by both messages simultaneously at any particular time. This may be crucial in the IoBT setting where the devices require information of both layers to make accurate decisions. Note that the developed dynamics can be easily extended to the case of multiple message propagation, which will require a larger state space; however, in this paper, we restrict ourselves to the case of only two network layers and messages.

\subsection{Steady State Analysis}
\textcolor{black}{We are interested in determining the steady state or equilibrium of the information dissemination process since it characterizes the eventual spread of information in the network which is independent of time. Although, with the changes in network topology and other network configurations, the actual information spread might be different; however, the equilibrium state provides us with a reasonable understanding of the system behavior.} For the single message dissemination case, we impose the stationarity condition, i.e., set $\frac{d I^{(c)}_{k}(t)}{dt} = 0$. It results in the following expression:
\begin{align} \label{I_k_expression}
I^{(c)*}_k = \frac{\alpha^{(c)} k \Theta^{(c)}(\alpha^{(c)})}{1 + \alpha^{(c)} k \Theta^{(c)}(\alpha^{(c)})}.
\end{align}
Notice that $\Theta^{(c)}(\alpha^{(c)})$ is now a constant that depends on $\alpha^{(c)}$. Now, \eqref{theta_expression} and \eqref{I_k_expression} present a system of equations that needs to be solved self-consistently to obtain the solution for $\Theta^{(c)}(\alpha^{(c)})$ and $I^{(c)*}_k$. For the multiple message propagation case, we can first write the reduced system in terms of the three independent states by substituting $UU_{k,l}(t) = 1 - IU_{k,l}(t) - UI_{k,l}(t) - II_{k,l}(t)$ to obtain \eqref{diff1_rev}, \eqref{diff2_rev}, and \eqref{diff3_rev}.

\begin{figure*}
\begin{align}
\frac{dIU_{k,l}(t)}{dt} &= \alpha^{(1)} k \Theta_1 - (\alpha^{(1)} k \Theta_1 + \alpha^{(2)} l \Theta_2 + 1)IU_{k,l}(t) - \alpha^{(1)} k \Theta_1 UI_{k,l}(t) - (\alpha^{(1)} k \Theta_1 - 1) II_{k,l}(t), \label{diff1_rev} \\
\frac{dUI_{k,l}(t)}{dt} &= \alpha^{(2)} l \Theta_2 - \alpha^{(2)} l \Theta_2 IU_{k,l}(t) + (\alpha^{(1)} k \Theta_1 + \alpha^{(2)} l \Theta_2 + 1)UI_{k,l}(t) - (\alpha^{(2)} l \Theta_2 - 1)II_{k,l}(t), \label{diff2_rev} \\
\frac{dII_{k,l}(t)}{dt} &= \alpha^{(1)} k \Theta_1 UI_{k,l}(t) + \alpha^{(2)}  l \Theta_2 IU_{k,l}(t) - 2II_{k,l}(t). \label{diff3_rev}
\end{align}
\end{figure*}
Again, using the stationarity condition, i.e., $\frac{dIU_{k,l}(t)}{dt} = 0$, $\frac{dUI_{k,l}(t)}{dt} = 0$, and $\frac{dII_{k,l}(t)}{dt} = 0$, we obtain the following expressions:
\begin{align}
IU_{k,l}^{*} = \frac{\alpha^{(1)} k \Theta_1}{(1 + \alpha^{(1)} k \Theta_1)(1 + \alpha^{(2)} l \Theta_2)}, \label{self1}\\
UI_{k,l}^{*} = \frac{\alpha^{(2)} l \Theta_2}{(1 + \alpha^{(1)} k \Theta_1)(1 + \alpha^{(2)} l \Theta_2)}, \label{self2}\\
II_{k,l}^{*} = \left(\frac{\alpha^{(1)} k \Theta_1}{1 + \alpha^{(1)} k \Theta_1} \right) \left(\frac{\alpha^{(2)} l \Theta_2}{1 + \alpha^{(2)} l \Theta_2} \right), \label{self3}
\end{align}

Now, the equations \cref{self1,self2,self3} and  \cref{eq_sigma1,eq_sigma2} need to be solved self-consistently to obtain the equilibrium solution of the quantity of interest, i.e., $II_{k,l}^{*}$.
In the subsequent section, we present the methodology to obtain the solution to the dynamical information spreading process for the IoBT network and develop a framework that can assist in the planning and efficient design of such networks.

\section{Methodology}
In this section, we first present a solution to the dynamical information spreading system developed in Section~\ref{sec:dynamics} and then use it for the efficient design of IoBT networks for mission-specific battlefield applications.

\subsection{Equilibrium Solution}
In order to find the equilibrium solution for the single network-wide message propagation case, we need to solve the self-consistent system expressed in \eqref{theta_expression} and \eqref{I_k_expression}. In fact, it reduces to obtaining a solution to the following fixed-point system:
\begin{align} \label{fixed-point-equation}
\Theta^{(c)}(\alpha^{(c)}) = \frac{1}{\mathbb{E}[K_c]} \sum_{k\geq 0} k \mathbb{P}(K_c = k) \frac{\alpha^{(c)} k \Theta^{(c)}(\alpha^{(c)})}{1 + \alpha^{(c)} k \Theta^{(c)}(\alpha^{(c)})}.
\end{align}
An obvious solution for this fixed-point system is $\Theta^{(c)}(\alpha^{(c)})= 0$; however, it is noninformative. The existence of a nonzero solution is stated in the following theorem:
\begin{theorem} \label{ex_and_uniq}
The fixed point equation in \eqref{fixed-point-equation} relating to the information dissemination dynamics may have at least one solution in the domain $\Theta^{(c)}(\alpha^{(c)}) > 0$ depending on the value of $\alpha^{(c)}$. The condition for this bifurcation to hold is $\alpha^{(c)} \geq \frac{\mathbb{E}[K_c]}{\mathbb{E}[K_c^2]}$. This bifurcation point is unique in the domain $0 < \Theta^{(c)}(\alpha^{(c)}) \leq 1$.
\begin{proof*}
See \textbf{Appendix~\ref{existence}}.
\end{proof*}
\end{theorem}
Obtaining a closed form solution for the fixed point system in~\eqref{fixed-point-equation} for a PPP setting is not possible due to the complex form of $\mathbb{P}(K_c = k)$. Hence, an approximate solution can be obtained using the following theorem:
\begin{theorem} \label{theorem_solution}
If a nonzero solution exists for the information spreading dynamics in \eqref{theta_expression} and \eqref{I_k_expression}, i.e., $\alpha^{(c)} \geq \frac{\mathbb{E}[K_c]}{\mathbb{E}[K_c^2]}$, then for $\mathbb{E}[K_c] \gg 1$, a lower bound approximation of the solution can be expressed as follows:
{
\begin{align}
\Theta^{(c)}(\alpha^{(c)}) \approx \left(1 - \frac{1}{\alpha^{(c)} \mathbb{E}[K_c]}\right)^+, \
\end{align}}
where $(x)^+$ represents $\max(0,x)$.
\begin{proof*}
See \textbf{Appendix~\ref{solution}}.
\end{proof*}
\end{theorem}
As shown in Appendix~\ref{solution}, Theorem~\ref{theorem_solution} provides a lower bound for the exact solution and becomes a tight approximation for $\mathbb{E}[K_c]~\gg~1$. Moreover, the approximation has also resulted in an increase of the critical information spreading threshold to $\alpha^{(c)} \geq \frac{1}{\mathbb{E}[K_c]}$ to ensure that $\Theta^{(c)} \geq 0$. This is also asymptotically accurate as the original condition can be written as $\alpha^{(c)} \geq \frac{\mathbb{E}[K_c]}{\mathbb{E}[K_c^2]} = \frac{1}{\mathbb{E}[K_c] + \frac{\sigma^2_{K_c}}{\mathbb{E}[K_c]}}$, where $\sigma^2_{K_c}$ is the variance of $K_c$. It approaches $\frac{1}{\mathbb{E}[K_c]}$ as $\mathbb{E}[K_c]$ becomes large. The relaxed condition is formally expressed by the following corollary.
\begin{corollary}\label{cond_approx}
The condition for obtaining an approximate nonzero equilibrium for the information dissemination dynamics is given as follows:
{
\begin{align}
\alpha^{(c)} \geq \frac{1}{\mathbb{E}[K_c]}.
\end{align}}
\end{corollary}
In the IoBT network, the physical interpretation of $\mathbb{E}[K_c]$ is that the average number of communication neighbours of a device in the combined network and hence, it is reasonable to assume that $\mathbb{E}[K_c] \geq 1$ due to the potential high density of devices in IoBT networks. Therefore, the solution presented in Theorem~\ref{theorem_solution} is indeed a good approximation to the actual solution. The corresponding solution for the density of informed devices can be obtained using~\eqref{I_k_expression}.

In the case of multiple message propagation, we need to solve the self-consistent system of equations defined by the equations \cref{self1,self2,self3} and  \cref{eq_sigma1,eq_sigma2}. It reduces to solving the fixed-point equations given in \eqref{fp1} and \eqref{fp2}.

\begin{figure*}
\begin{align}
&\Theta_1 = \frac{1}{\mathbb{E}[K_1]} \sum_{k,l} \mathbb{P}(K_1 = k,K_2 = l) k \left( \frac{\alpha^{(1)} k \Theta_1 + {\alpha^{(1)} }^2 kl \Theta_1 \Theta_2}{(1 + \alpha^{(1)} k \Theta_1)(1 + \alpha^{(2)} l \Theta_2)} \right) = \frac{1}{\mathbb{E}[K_1]} \sum_{k} \mathbb{P}(K_1 = k) k \left( \frac{\alpha^{(1)} k \Theta_1}{1 + \alpha^{(1)} k \Theta_1} \right), \label{fp1}\\
&\Theta_2 = \frac{1}{\mathbb{E}[K_2]} \sum_{k,l} \mathbb{P}(K_1 = k,K_2 = l) l \left( \frac{\alpha^{(2)} l \Theta_2 + {\alpha^{(2)}}^2 kl \Theta_1 \Theta_2}{(1 + \alpha^{(1)} k \Theta_1)(1 + \alpha^{(2)} l \Theta_2)} \right) = \frac{1}{\mathbb{E}[K_2]} \sum_{l} \mathbb{P}(K_2 = l) l \left( \frac{\alpha^{(2)} l \Theta_2}{1 + \alpha^{(2)} l \Theta_2} \right). \label{fp2}
\end{align}
\end{figure*}
Notice that $\Theta_1$ and $\Theta_2$ are independent of each other and their fixed point equations are similar to \eqref{fixed-point-equation}. Hence, the existence and uniqueness of the fixed point can be proved under similar conditions. Using a similar approach to the solution of the single message propagation case, a lower bound approximate solution is provided by the following theorem:
\begin{theorem} \label{thm2}
If a nonzero solution exists for the information spreading dynamics for the case two simultaneous message propagation, i.e., \textcolor{black}{$\alpha^{(1)} \geq  \frac{\mathbb{E}[K_1]}{\mathbb{E}[K_1^2]}$ and $\alpha^{(2)} \geq \frac{\mathbb{E}[K_2]}{\mathbb{E}[K_2^2]}$, then for $\mathbb{E}[K_1] \gg 1$ and $\mathbb{E}[K_2] \gg 1$}, a lower bound approximation of the solution can be expressed as follows:
{
\begin{align}
\Theta_1(\alpha)  \approx \left(1 - \frac{1}{\alpha^{(1)} \mathbb{E}[K_1]} \right)^+,\\
\Theta_2(\alpha)  \approx \left( 1 - \frac{1}{\alpha^{(2)} \mathbb{E}[K_2]} \right)^+.
\end{align}}
\textbf{\emph{ Proof.}} See \textbf{Appendix~\ref{thm2_proof}}.
\end{theorem}
Similar to Corollary~\ref{cond_approx}, the condition for obtaining an approximate nonzero equilibrium solution is $\alpha^{(i)} \geq 1/\mathbb{E}[K_i], i \in \{1,2\}$. The corresponding solution for the proportion of devices that are informed both messages, i.e., $II_{k,l}$ can be obtained using \eqref{self3} which turns out to be the following:
{
\begin{align} \label{decomposability}
II_{k,l}^* = I_k^{(1)*} \times I_l^{(2)*}.
\end{align}}
This result is interesting and useful as it can be easily generalized for the case of multiple connectivity layers, i.e., types of devices, and multiple messages propagating simultaneously.


\subsection{Secure and Reconfigurable Network Design}

\begin{figure}[t]
  \centering
  \includegraphics[width=3in]{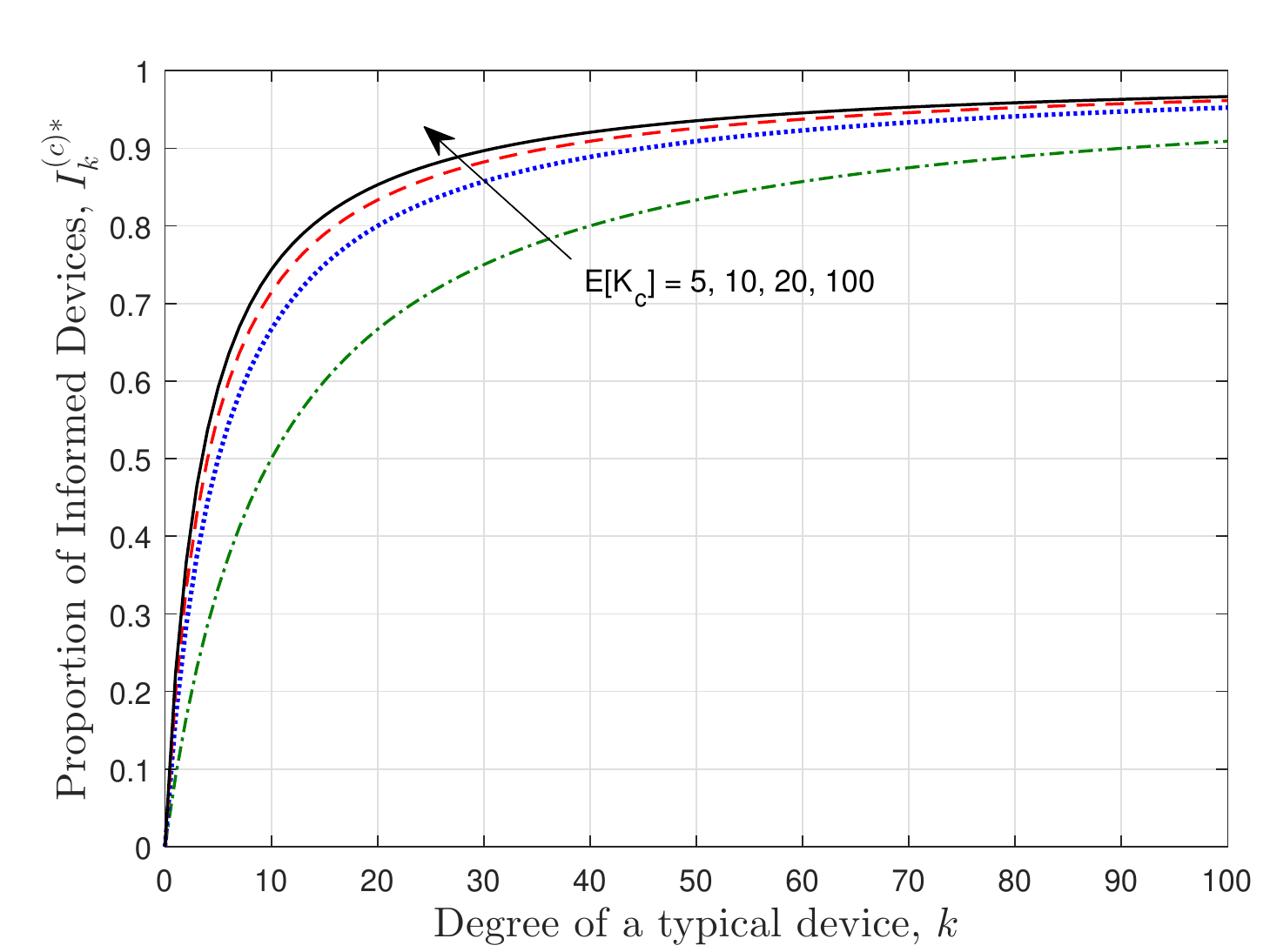}\\
  \caption{Information dissemination profiles for varying average degree of devices ($\alpha = 0.3$).}\label{trajectory}
\end{figure}

Once the equilibrium point for information dissemination has been determined, the next step is to design the IoBT network to achieve mission specific goals while efficiently using battlefield resources. In essence, the network design implies tuning the knobs of the network, which in the case of IoBT networks are the transmission ranges and the node deployment densities of the different types of battlefield things. However, changing the physical parameters may have an impact on the cost and hence, the goal is to ensure a certain information spreading profile in the network while deploying the minimum number of devices and using the minimum transmit power. Let $\mathbf{r} = [r_1 \ r_2]^T$ represent the vector of communication ranges of each of the type-$\RN{1}$ and type-$\RN{2}$ devices in the IoBT network. The minimum density of the devices in the network, determined by the mission requirements, is denoted by $\lambda^{\min}$, $\lambda^{\min} \geq 0$. The maximum deployment density of the devices, defined by the capacity of the available devices, is denoted by $\lambda^{\max}$, $\lambda^{\max} \geq \lambda^{\min}$. Similarly, the tunable transmission range limits of the devices can be expressed as $\mathbf{r}^{\min} = [r_1^{\min} \ r_2^{\min}]^T$, $r_m^{\min} \geq 0, \ \forall m \in \{1,2 \}$, and $\mathbf{r}^{\max} = [r_1^{\max} \ r_2^{\max}]^T$, $r_m^{\max} \geq r_m^{\min}, \ m \in \{ 1,2 \}$. If $\mathbf{w} = [w_1 \ w_2]^T$ such that $\sum_{m=1}^{2} w_m = 1$ represents the weight vector corresponding to the relative capital cost of deploying a type-$\RN{1}$ and type-$\RN{2}$ device respectively, and $c$ represents the unit operational power cost signifying the importance of network power consumption, then a cost function for the network with device densities $\boldsymbol{\lambda}$ and transmission ranges $\mathbf{r}$ can be expressed as follows:
\begin{align}
\mathcal{C}(p,\lambda,\mathbf{r}) = w_1 p \lambda + w_2 (1-p)\lambda + c (p \lambda r_1^{\eta} + \lambda r_2^{\eta}),
\end{align}
where $\eta$ denotes the path-loss exponent\footnote{The power consumption of a device with communication range $r_m$ is proportional to $r_m^{\eta}$.}. The first term represents the total deployment cost per unit area of all the network devices while the second term represents the total energy cost per unit area of operating all the devices with transmission range $\mathbf{r}$. The weights $\mathbf{w}$ can depend on several factors such as the time required for deployment, the monetary cost involved, or the number of devices available in stock, etc. We can then formulate the secure and reconfigurable network design problem as follows:
\begin{align}
& \underset{p, \lambda,\mathbf{r}}{\text{minimize}}
& & \mathcal{C}(p, \lambda,\mathbf{r}) \label{org_objective}\\
& \text{subject to}
& & I_k^{(c)*} \geq T^{(c)}_k, \forall k \geq 0, \label{org_const1}\\
&&& II_{k,l}^{*} \geq T_{k,l}, \forall k \geq 0, l \geq 0, \label{org_const2}\\
&&& p^{\min} \leq p \leq p^{\max}, \lambda^{\min} \leq \lambda \leq \lambda^{\max}, \notag \\ &&& \mathbf{r}^{\min} \leq \mathbf{r} \leq \mathbf{r}^{\max},\label{org_const4}
\end{align}
where $T^{(c)}_k \in (0,1)$, $k \geq 0$ are the minimum desired proportions of degree $k$ devices that are informed with a single network-wide message, $T_{k,l} \in (0,1)$, $k \geq 0, l \geq 0$ are the desired proportion of devices with degree $K_1 = k$ and $K_2 = l$ that are informed with both messages simultaneously, and $p^{\min}$ and $p^{\max}$ are the minimum and maximum fractions of devices respectively that are of type-$\RN{1}$. The cost function $\mathcal{C}$ is a convex nondecreasing function of $p$, $\lambda$, and $\mathbf{r}$. However, it is an infinite dimension optimization problem due to a constraint on each degree class of devices. To be able to solve this problem, we need to select a desired mapping $T^{(c)}_k$ for the combined degree $K_c = k$ and similarly a mapping $T_{k,l}$ for devices with joint intra-layer degrees $K_1 = k, K_2 = l$, $\forall k,l \geq 0$. Further investigation reveals that for a fixed $\alpha$, the information dissemination can only have restricted trajectories based on the average device degree as shown in Fig.~\ref{trajectory}. Hence, the threshold mappings cannot be defined arbitrarily as they might not be achievable. To handle this  problem, we can express the constraints in~\eqref{org_const1}, assuming that $\alpha^{(c)}$ satisfies the condition in Corollary~\ref{cond_approx}, as follows:
\begin{align} \label{modified_const_1}
\mathbb{E}[K_c] \geq \frac{1}{\alpha^{(c)} - T^{(c)}_k / \left(k (1 - T^{(c)}_k)\right)}, \forall k \geq 0,
\end{align}
Now, we need to satisfy~\eqref{modified_const_1} for each ($k, T_k^{(c)}$) pair, and from Fig.~\ref{trajectory}, it is
clear that there is no incentive to choose a threshold profile that is different from one of the possible trajectories. Therefore, specifying the threshold for a single value of the degree is sufficient to characterize the entire trajectory. Since $K_c$ is a random variable with a distribution centered around $\mathbb{E}[K_c]$, it is plausible to set a threshold on the proportion of devices with the average device degree, i.e., using the pair ($\mathbb{E}[K_c], T^{(c)}_{\mathbb{E}[K_c]}$). It results in the following single constraint instead of the infinite set of constraints in~\eqref{org_const1}:
\begin{align}
\mathbb{E}[K_c] \geq \frac{1}{\alpha^{(c)} (1 - T^{(c)}_{\mathbb{E}[K_c]})}.
\end{align}
Note that this constraint implies $\alpha^{(c)} \mathbb{E}[K_c] \geq 1/(1 - T^{(c)}_{\mathbb{E}[K_c]})$, which satisfies the condition in Corollary~\ref{cond_approx}, $ \forall\  T^{(c)}_{\mathbb{E}[K_c]} \in [0,1]$, thus validating our assumption.
For the second set of constraints in~\eqref{org_const2}, due to the decomposability of $II^*_{k,l}$ as shown in~\eqref{decomposability} and the fact that $I^{(i)*}_k \in [0,1], i \in \{1,2\}$, the constraints can be separated as follows:
\begin{align}
\frac{\alpha^{(1)} k \Theta_1}{1 + \alpha^{(1)} k \Theta_1} \geq T_{k,l}, \ \frac{\alpha^{(2)} l \Theta_2}{1 + \alpha^{(2)} l \Theta_2} \geq T_{k,l}, \forall k,l \geq 0.
\end{align}
Moreover, the thresholds can be replaced by simply $T_k$ and $T_l$ instead of $T_{k,l}$ as they are constants. Using a similar approach as before, we replace the set of infinite constraints in~\eqref{org_const2} by the following two constraints:
\begin{align}
\mathbb{E}[K_i] \geq \frac{1}{\alpha^{(i)} (1 - T_{\mathbb{E}[K_i]})}, i \in \{1,2\}.
\end{align}
For brevity, we henceforth denote $T^{(c)}_{\mathbb{E}[K_c]}$, $T_{\mathbb{E}[K_1]}$, and $T_{\mathbb{E}[K_2]}$ as simply $T^{(c)}$, $T^{(1)}$, and $T^{(2)}$. \textcolor{black}{Furthermore, we denote $\frac{1}{\alpha^{(c)} (1 - T^{(c)})}$, $\frac{1}{\alpha^{(1)} (1 - T^{(1)})}$, and $\frac{1}{\alpha^{(2)} (1 - T^{(2)})}$ by $\mathcal{T}^{(c)}$, $\mathcal{T}^{(1)}$, and $\mathcal{T}^{(2)}$ respectively corresponding to the desired minimum success probabilities $\mathcal{P}_s^{(i)}$ leading to $\alpha^{i}, i \in \{1,2,c\}$.} The original optimization problem can then be rewritten as follows:
\begin{align}
& \underset{p, \lambda,\mathbf{r}}{\text{minimize}}
& & w_1 p \lambda + w_2 (1-p)\lambda + c (p \lambda r_1^{\eta} + \lambda r_2^{\eta}), \label{mod_objective}\\
& \text{subject to}
& & p^2 \lambda \pi r_1^2 + \lambda \pi r_2^2 \geq \mathcal{T}^{(c)}, \label{mod_const1}\\
&&& p^2 \lambda \pi r_1^2 \geq \mathcal{T}^{(1)}, \label{mod_const2}\\
&&& \lambda \pi r_2^2 \geq \mathcal{T}^{(2)}, \label{mod_const3}\\
&&& p^{\min} \leq p \leq p^{\max}, \lambda^{\min} \leq \lambda \leq \lambda^{\max}, \notag \\ &&& \mathbf{r}^{\min} \leq \mathbf{r} \leq \mathbf{r}^{\max}.\label{mod_const4}
\end{align}
It is important to note that the conditions for existence of nonzero equilibrium in Corollary~\ref{cond_approx} and the ones resulting from Theorem~\ref{thm2} are implicitly incorporated into the constraints and do not need to be imposed separately. This implies that if a feasible solution to the optimization problem exists, then there exists a nonzero equilibrium solution to the information dissemination dynamics.
The objective and constraints are nondecreasing smooth functions of the optimization variables. \textcolor{black}{
It is clear that the objective and constraints are convex in the feasible solution space. Therefore, the problem can be solved using standard convex optimization techniques~\cite{convex_optimization}.
}

\begin{algorithm}
\begin{spacing}{1.2}
\caption{Secure and Reconfigurable Network Design}
\label{Algorithm1}
\begin{algorithmic}[1]
\STATE{At epoch, i.e., $t = 0$; Initialize requirements for information dissemination, i.e., $T^{(1)}$, $T^{(2)}$, and $T^{(c)}$ and the anticipated threat level $\delta$.}
\STATE{Obtain optimal network parameters $\lambda_1^{\text{init}} = p^{\text{init}} \lambda^{\text{init}}, \lambda_2^{\text{init}} = \lambda^{\text{init}}, r_1^{\text{init}}$, and $r_2^{\text{init}}$ by solving the optimization problem in~\cref{mod_objective,mod_const1,mod_const2,mod_const3,mod_const4} and accordingly deploy the devices with the appropriate communication ranges.}
\REPEAT \label{loop_start}
\IF{$t = \zeta t_{\text{r}}, \zeta \in \mathbb{Z}^+ $}
\STATE{Obtain an estimate of the density of active devices $\hat{\boldsymbol{\lambda}} = [\hat{\lambda}_{1},\hat{\lambda}_2]$ and use the initial communication ranges to estimate the prevailing information dissemination level $\hat{T}^{(1)}$, $\hat{T}^{(2)}$, and $\hat{T}^{(c)}$}.
\STATE{Re-evaluate the desired security level in response to the threats and accordingly update the parameter $\hat{\delta}$.}
\IF{$|T^{(1)} - \hat{T}^{(1)}| \geq \epsilon$ \textbf{or} $|T^{(2)} - \hat{T}^{(2)}| \geq \epsilon$ \textbf{or} $|T^{(c)} - \hat{T}^{(c)}| \geq \epsilon$ \textbf{or} $\hat{\delta} \neq \delta$,}
\STATE{Recompute the optimization problem in~\cref{mod_objective,mod_const1,mod_const2,mod_const3,mod_const4} to obtain the new set of optimal parameters $\lambda_1^{\text{new}} \gets p^{\text{new}} \lambda^{\text{new}}$, $\lambda_2^{\text{new}} \gets \lambda^{\text{new}}$, $r_1^{\text{new}}$, and $r_2^{\text{new}}$.}
\STATE{Deploy additional required $\boldsymbol{\lambda}^{\text{new}} - \hat{\boldsymbol{\lambda}}$ devices in the network and reconfigure transmission powers to achieve the required transmission ranges $\mathbf{r}^{\text{new}}$.}
\ENDIF
\ENDIF
\STATE $t \gets t + 1$;
\UNTIL{End of mission.} \label{loop_end}
\end{algorithmic}
\end{spacing}
\end{algorithm}

The secure and resilient framework for mission critical information dissemination in IoBT networks is provided in Algorithm~\ref{Algorithm1}. At the beginning of the mission, the central network planner obtains the optimal physical network parameters by solving the optimization problem in~\cref{mod_objective,mod_const1,mod_const2,mod_const3,mod_const4}, and accordingly deploys the devices with appropriate transmission powers. The central planner then periodically analyzes the connectivity situation of the network with a reconfigurability interval denoted by $t_r$. The reconfigurability interval could range from several hours to days depending on the mission requirements. Several techniques may be employed to estimate the connectivity, and consequently the information dissemination in the network. The effect of physical damage can be measured by physically monitoring the network with the aid of robotic systems such as in~\cite{density_estimation}. On the other hand, the effect of cyber attacks may be estimated by running discovery tests on the network to assess the reachability of nodes. Based on these results, estimates of the information dissemination thresholds can be obtained, i.e., $\hat{T}^{(1)}$, $\hat{T}^{(2)}$, and $\hat{T}^{(c)}$. If there is a significant drop in the estimated threshold as compared to the desired one or there is a change in the required security level, then the optimization needs to be re-computed. The new optimized parameters help in identifying the additional deployment needed for each type of devices and/or the reconfiguration of their transmission powers to achieve the desired information dissemination in the IoBT network. In the following section, we investigate the behaviour of the optimal solutions under varying threat levels and mission specific performance thresholds.

\section{Results}
In this section, we provide the results obtained by testing the developed framework under different battlefield missions. We assume a bi-layer IoBT network comprising of type-$\RN{1}$ and type-$\RN{2}$ devices. The first type of devices is assumed to be commanders and the second type is assumed to be followers. The assumption yields a simple yet natural network configuration in a battlefield, e.g., being composed of soldiers and distributed commanding units. The allowable physical parameter ranges of the respective devices are selected to be as follows: the minimum and maximum device deployment density is selected as $\lambda^{\min} = 1$ km$^{-2}$ and $\lambda^{\max} = 15$ km$^{-2}$ respectively, the minimum and maximum fraction of type-$\RN{1}$ devices $p^{\min} = 0$ and $p^{\max} = 0.4$ respectively, the minimum and maximum communication ranges of devices in the first layer, $r_1^{\min} = 100$ m and $r_1^{\max} = 2000$ m respectively, and the minimum and maximum communication ranges of devices in the second layer, $r_2^{\min} = 10$ m and $r_2^{\max} = 800$ m. The parameters imply that the active devices in the first layer, i.e., the commanding units have a higher allowable transmission range than the followers. In practice, the limits can be based on tactical requirements of the missions. \textcolor{black}{For simplicity, we assume that sufficient number of channels are available and for the considered densities and communication ranges of the devices, and the MAC protocol is able to effectively mitigate interference in the communication resulting in a constant success probability. Further, we assume that the desired success probabilities $\mathcal{P}_s^{(i)} = 1, \forall i \in \{1,2,c\}$ implying perfect success of transmissions. Consequently, $\alpha^{(1)} = \alpha^{(2)} = \alpha^{(c)}$ and is henceforth referred to as $\alpha$.} The weights representing the relative deployment cost are chosen to be $w_1 = 100$ and $w_2 = 50$, which implies that the deployment cost of the commanding units is twice as much as the follower units. The unit cost of power is selected to be $c = 100$ that can be adjusted according to the importance of each mission and the path-loss exponent $\eta = 4$.


\subsection{Mission Scenarios}
In the battlefields, there can be several types of missions such as intelligence, surveillance, encounter battle, espionage, reconnaissance, etc. In our results, we will focus particularly on the two most common mission scenarios, i.e., intelligence and encounter battle. Both of them have completely different requirements in terms of the information flow in the network, which are described as follows:

\subsubsection{Intelligence} 
\begin{figure*}[t!]
\addtolength{\subfigcapskip}{-0.0in}
\begin{center}
\subfigure[Transmission ranges of devices against varying threat level.]{\label{intel_range}\includegraphics[width=2.2in]{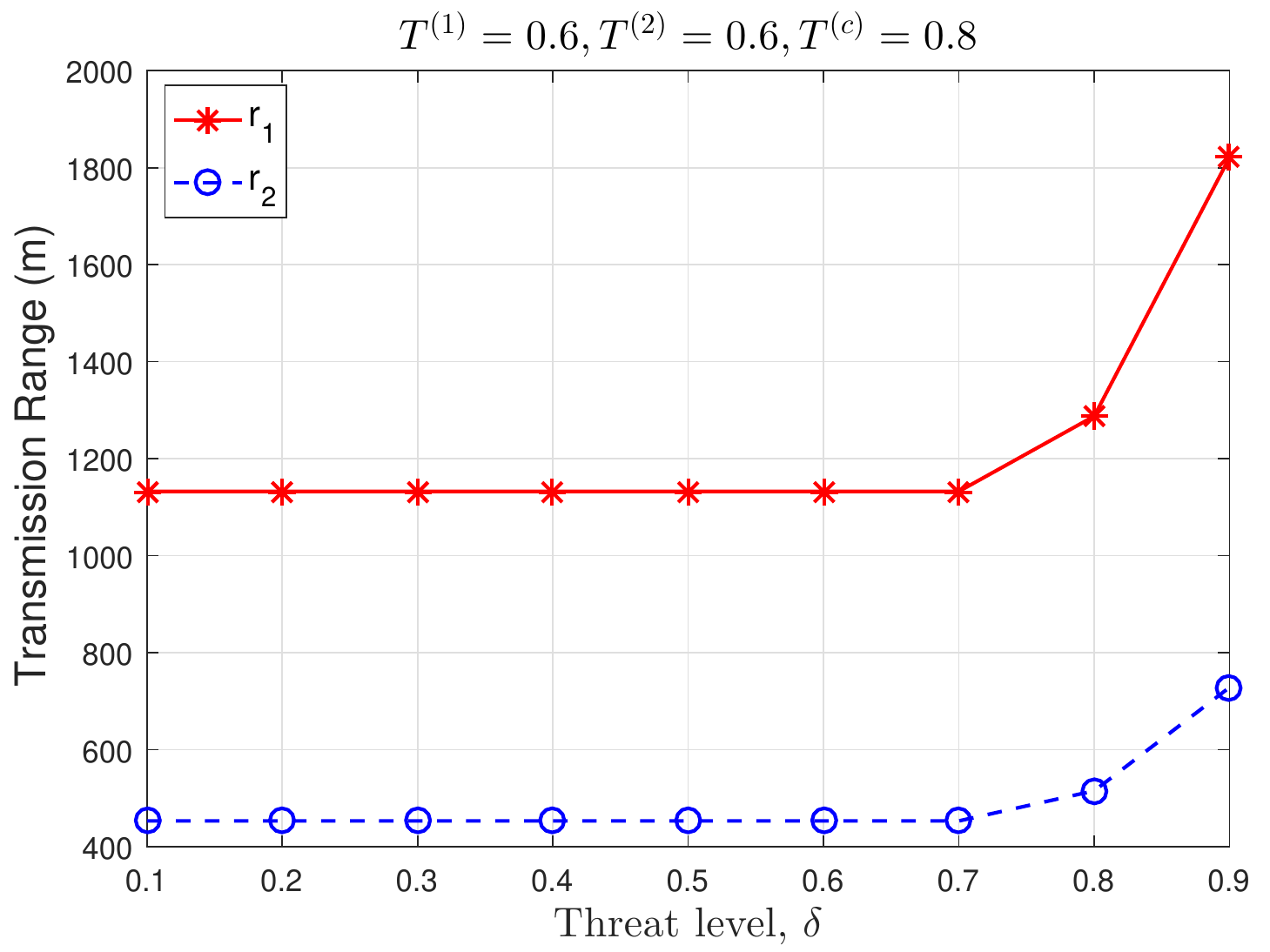}}
\subfigure[Deployment density of devices against varying threat level.]{\label{intel_density}\includegraphics[width=2.2in]{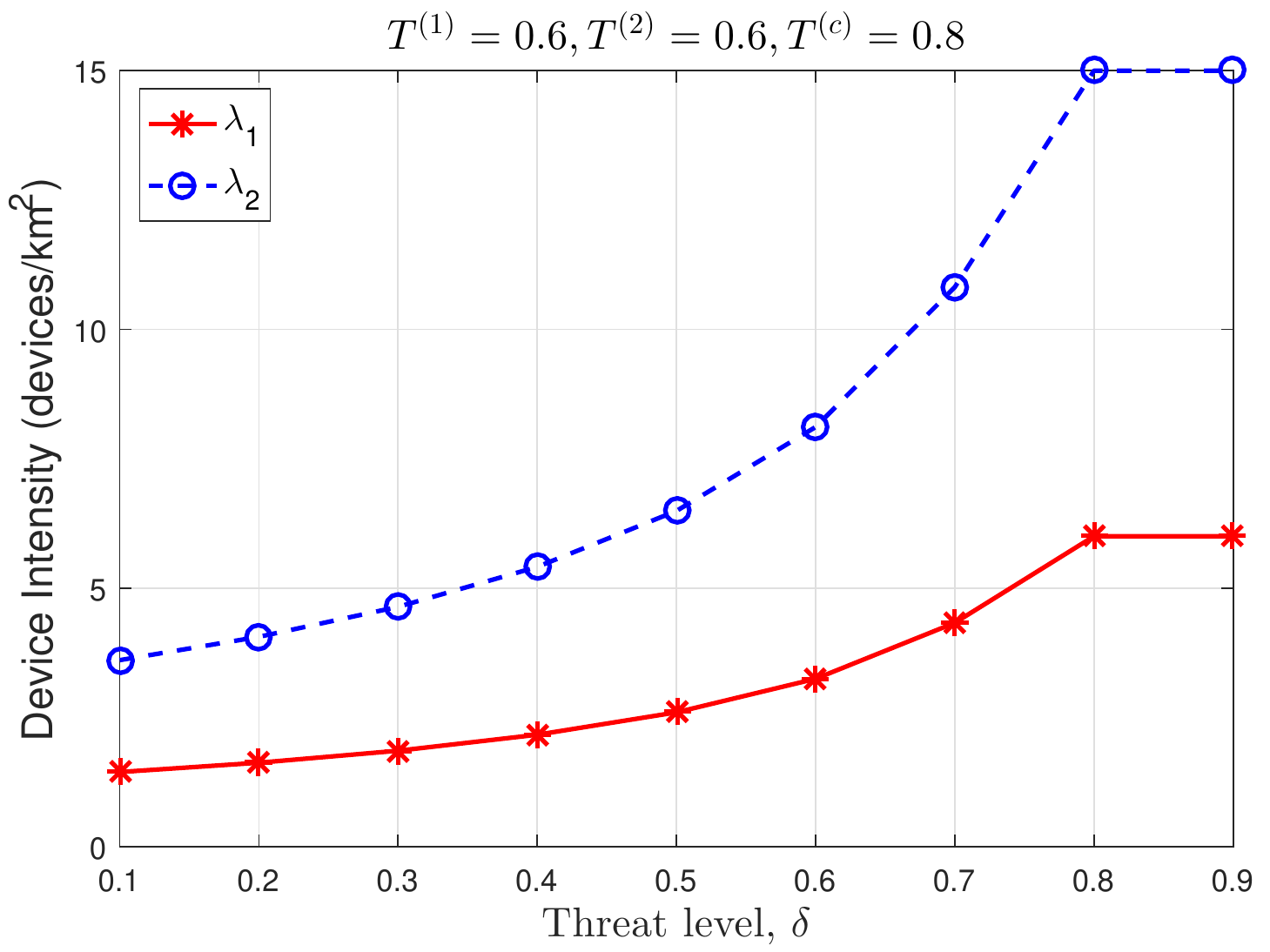}}\
\subfigure[Cost function against varying threat level.]{\label{intel_cost}\includegraphics[width=2.2in]{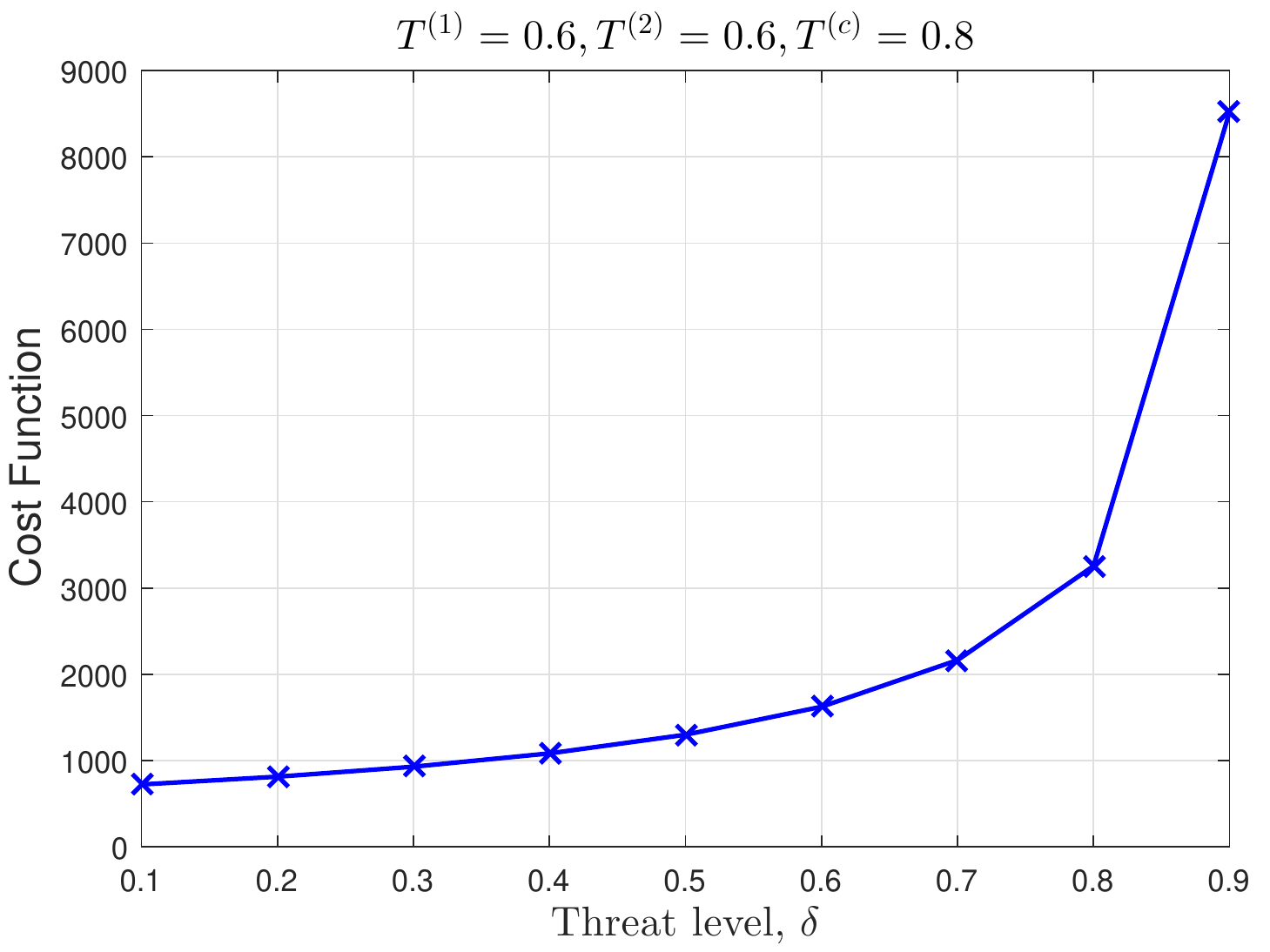}}
\end{center}
\caption{Optimal network parameters for the intelligence mission.}
\label{Fig_intelligence}
\end{figure*}

In the intelligence mission, the goal is to provide commanders with the information from a range of sources to assist them in operational or campaign planning. This requires a high network-wide dissemination of a single message, while the condition for simultaneously informed devices may not be stringent. In essence, the commander network must reinforce the follower network to achieve a high network-wide message propagation. Hence, to emulate such an intelligence mission, we select the following set of information spreading thresholds: $T^{(1)} = T^{(2)} = 0.6$, $T^{(c)} = 0.8$. The optimal physical parameters obtained for the intelligence mission against increasing threat level $\delta$ are shown in Fig.~\ref{Fig_intelligence}. There are several interesting observations in the intelligence mission. A general trend is that the required transmission ranges and deployment densities increases as the threat level increases. Consequently, the cost function, which signifies the deployment and operation cost of the network, also increases as shown in Fig.~\ref{intel_cost}. Fig.~\ref{intel_range} and Fig.~\ref{intel_density} show that the transmission range of the commanders is always higher than the followers while the densities of the followers is higher than that of the commanders. This observation makes sense as the followers equipped with sensors should be more in the total number while the commander network should have a larger influence area to be able to gather information for the intelligence mission. Another important observation is that the framework tends to increase the deployment density of the devices first before increasing their transmission ranges. It is due to a high cost of power consumption that tends to force the devices to minimize the transmission ranges.

\begin{figure*}[t!]
\addtolength{\subfigcapskip}{-0.0in}
\begin{center}
\subfigure[Transmission ranges of devices against varying threat level.]{\label{enc_range}\includegraphics[width=2.2in]{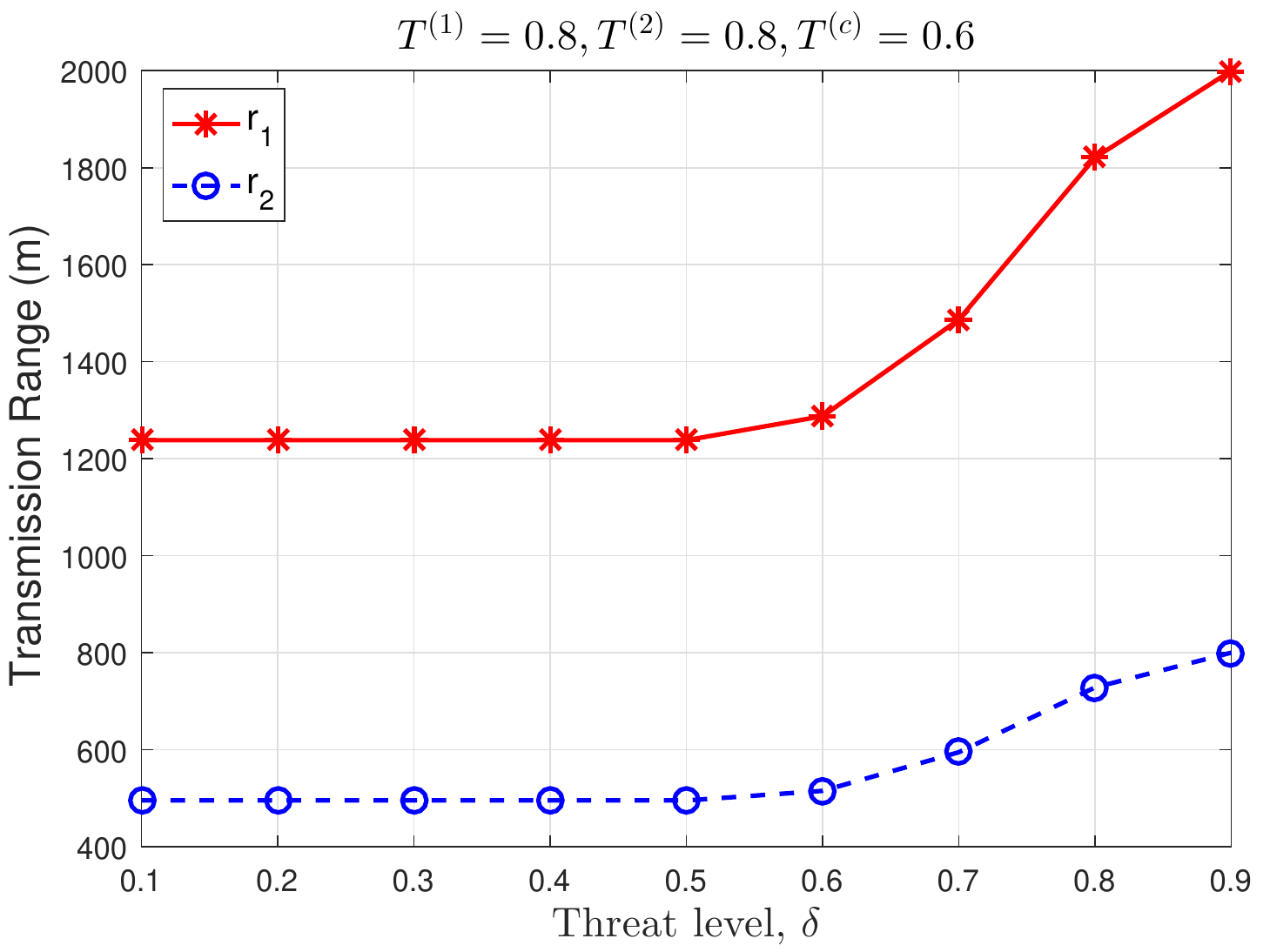}}
\subfigure[Deployment density of devices against varying threat level.]{\label{enc_density}\includegraphics[width=2.2in]{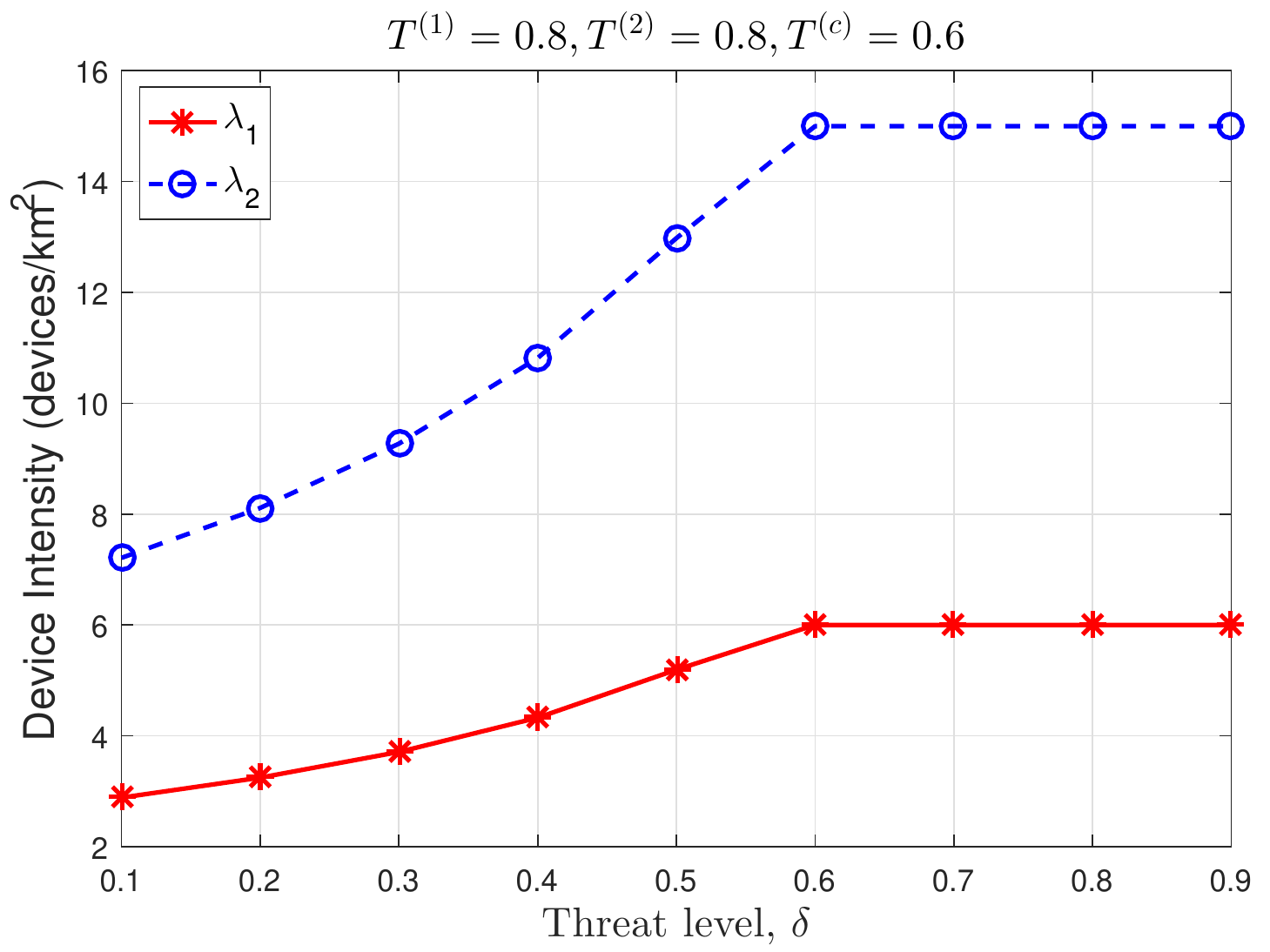}}\
\subfigure[Cost function against varying threat level.]{\label{enc_cost}\includegraphics[width=2.2in]{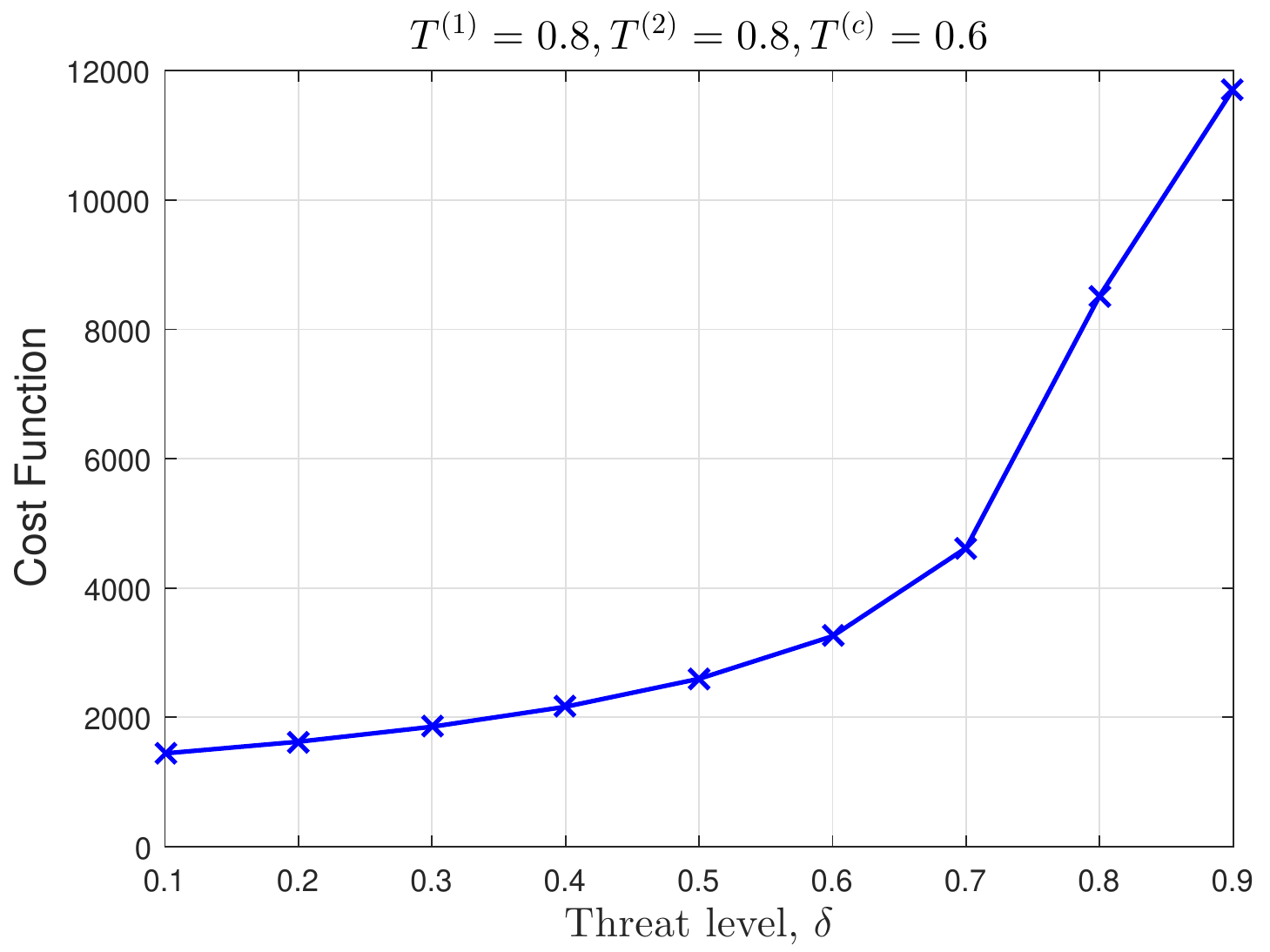}}
\end{center}
\caption{Optimal network parameters for the encounter mission.}
\label{Fig_encounter}
\end{figure*}

\subsubsection{Encounter Battle} In the encounter battle or meeting engagement scenario, there is a contact between the battling forces. In such situations, commanders need to act quickly to gain advantage over the opponents. This requires devices to be informed simultaneously about the information disseminated in both network layers. Hence, there is a need for strong coordination among commanders as well as followers. Additionally, the common status information sharing between all network devices must be sufficiently high to ensure accurate decision-making. Therefore, we set the following information spreading thresholds: $T^{(1)} = T^{(2)} = 0.8$ and $T^{(c)} = 0.6$. The resulting optimal parameters against the changing threat level are presented in Fig.~\ref{Fig_encounter}. In contrast with the intelligence mission, the framework utilizes all the available devices at a much lower threat level and is also forced to increase the transmission ranges to full capacity despite the high power cost.
The cost function for the encounter battle in Fig.~\ref{enc_cost} is significantly higher than the intelligence mission in Fig.~\ref{intel_cost} due to more stringent information spreading criteria which requires more transmission power and higher device density.

\begin{figure*}[t!]
\addtolength{\subfigcapskip}{-0.0in}
\begin{center}
\subfigure[Transmission ranges of devices against varying network-wide information dissemination threshold.]{\label{threshold_range}\includegraphics[width=2.5in]{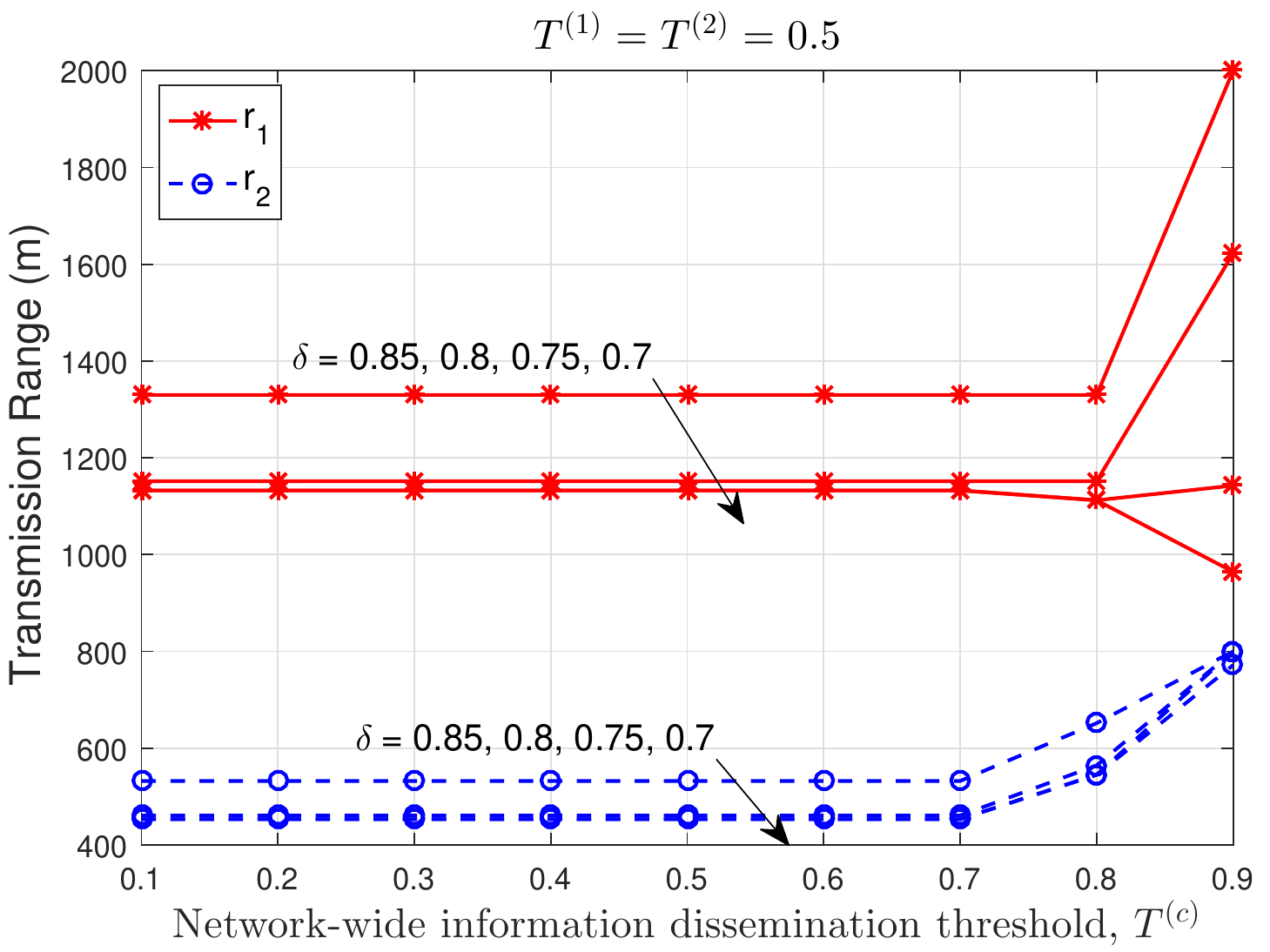}} \ \
\subfigure[Deployment density of devices against varying network-wide information dissemination threshold.]{\label{threshold_density}\includegraphics[width=2.5in]{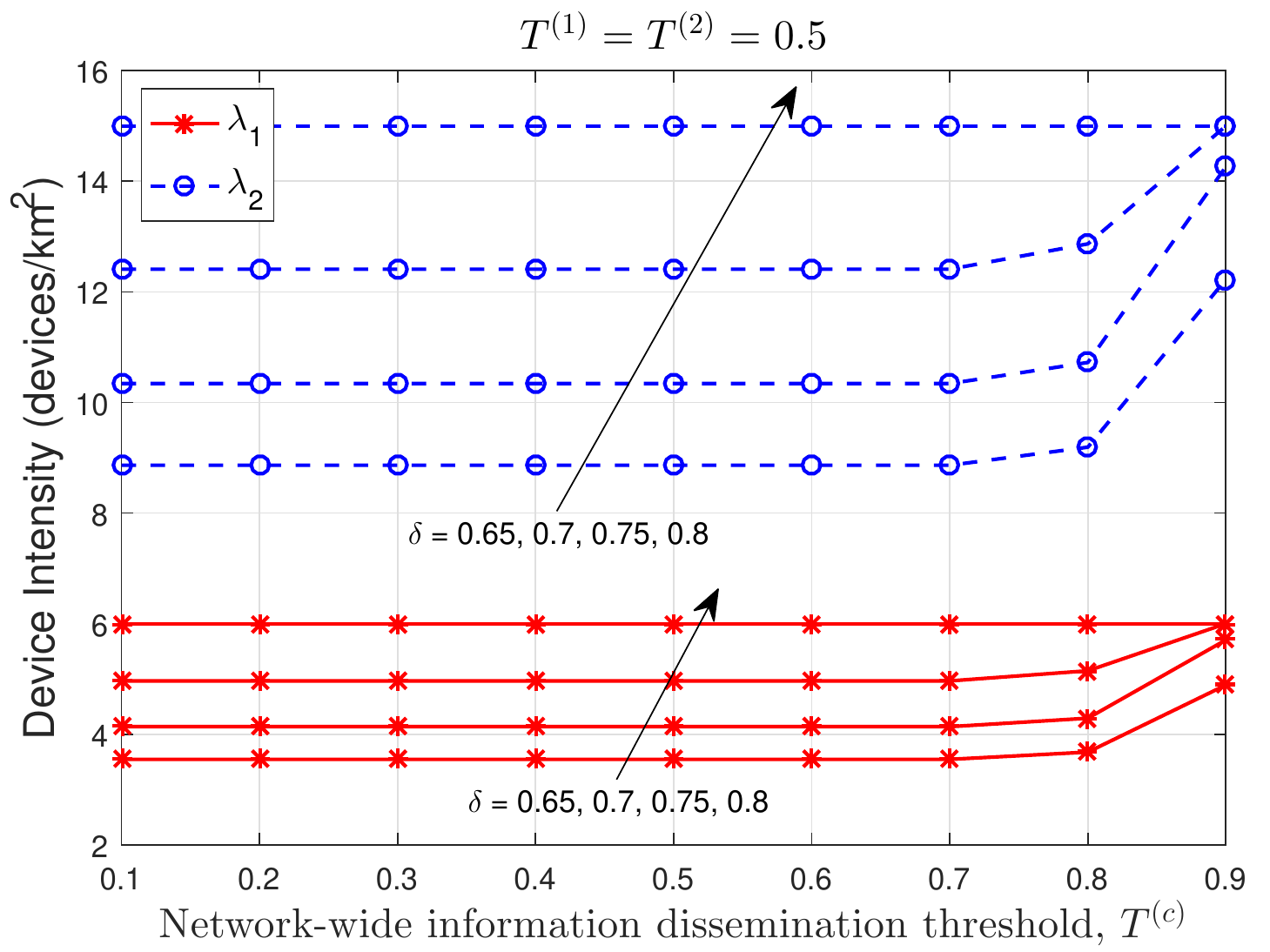}}\
\end{center}
\caption{Network parameter variation with changes in combined-layer information dissemination threshold.}
\label{Fig_threshold_combined}
\end{figure*}

\begin{figure*}[t]
\addtolength{\subfigcapskip}{-0.0in}
\begin{center}
\subfigure[Transmission ranges of devices against varying intra-layer information dissemination threshold.]{\label{threshold_range_individual}\includegraphics[width=2.5in]{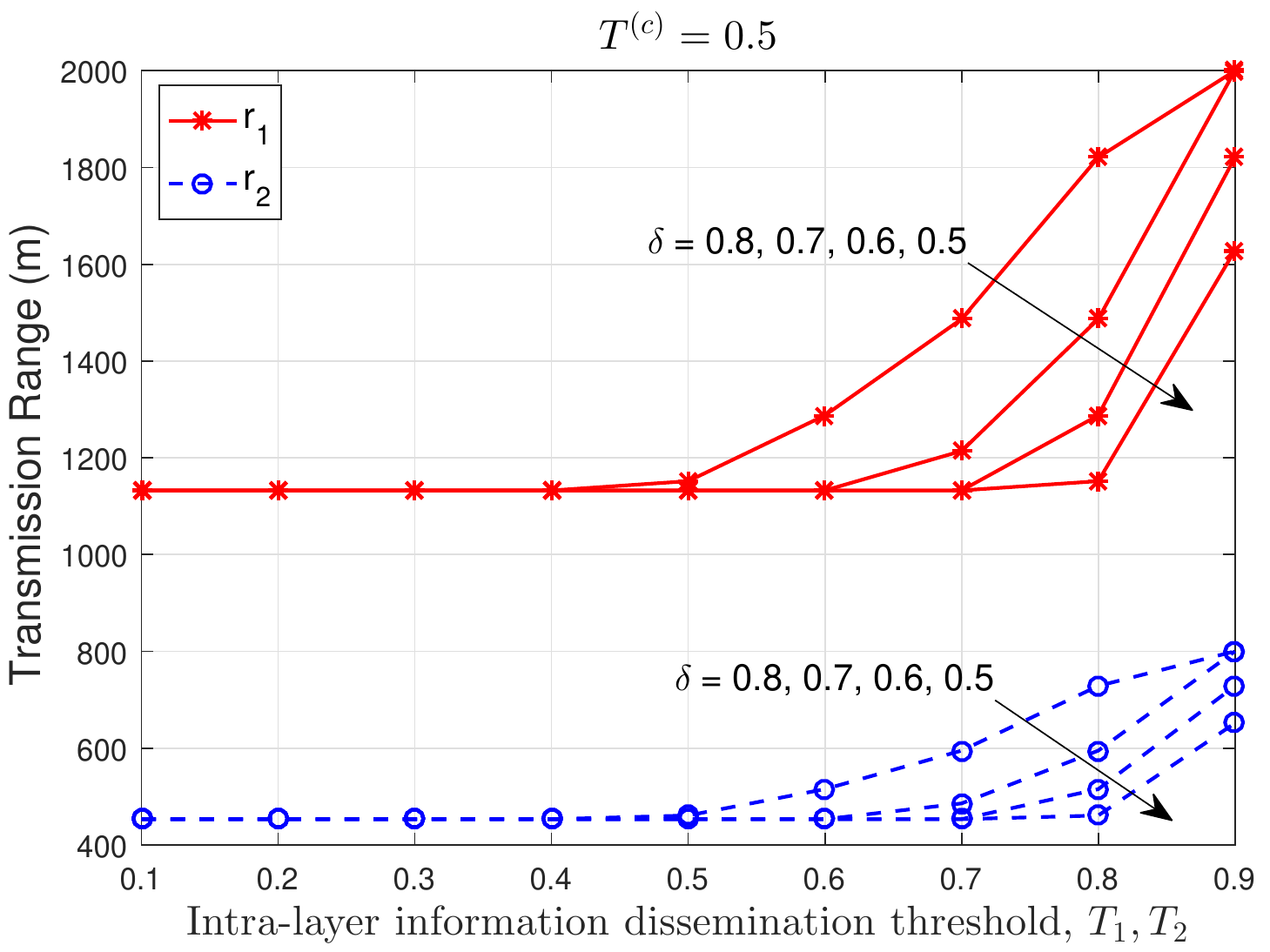}} \ \
\subfigure[Deployment density of devices against varying intra-layer information dissemination threshold.]{\label{threshold_density_individual}\includegraphics[width=2.5in]{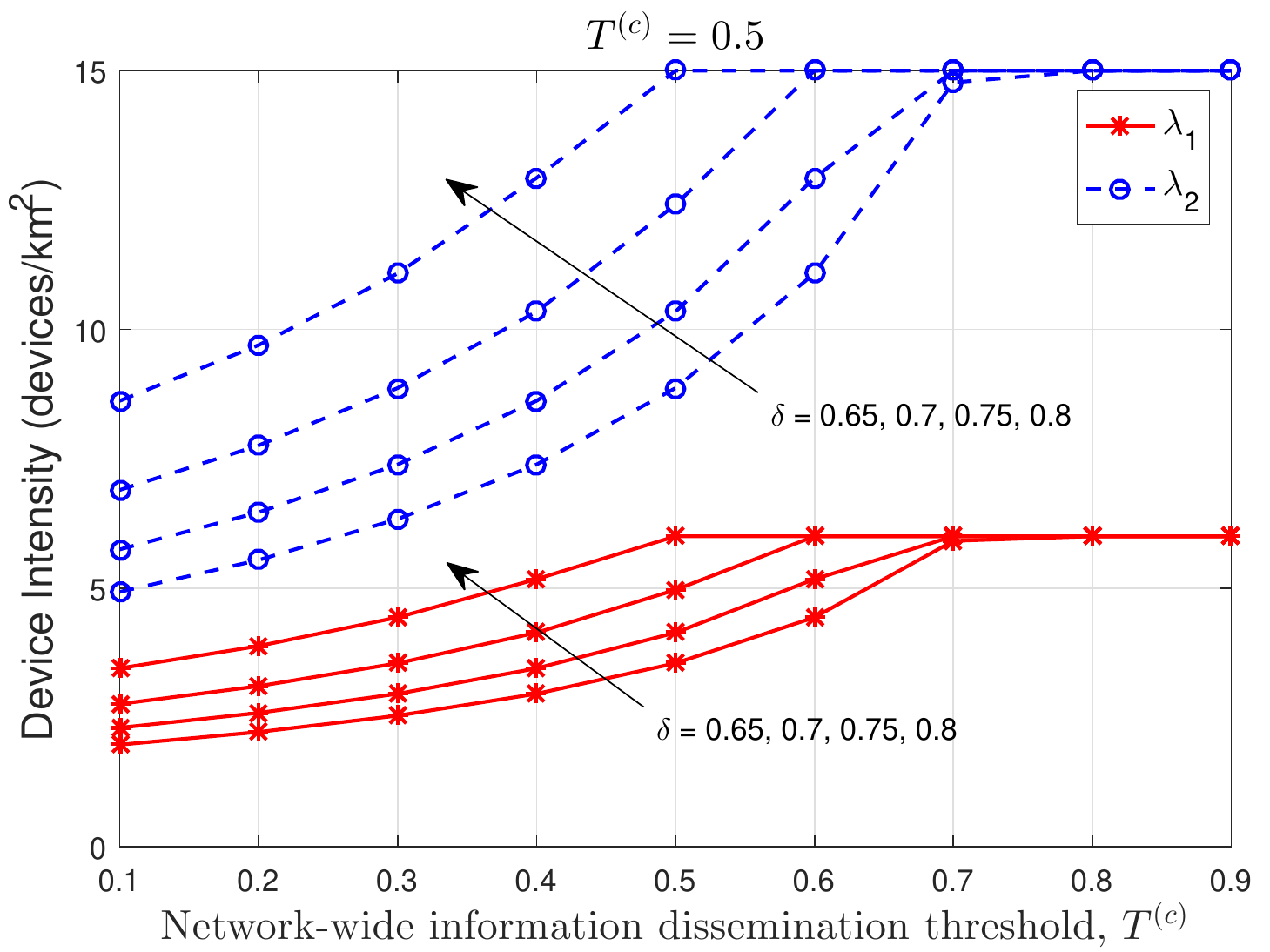}}\
\end{center}
\caption{Network parameter variation with changes in intra-layer information dissemination threshold.}
\label{Fig_threshold_individual}
\end{figure*}

Next, we study the impact of the information spreading thresholds on the optimal network parameters. In Fig.~\ref{Fig_threshold_combined}, we fix the intra-layer information dissemination threshold to $T^{(1)} = T^{(2)} = 0.5$ and observe the change in parameters for varying network-wide information dissemination threshold $T^{(c)}$. Notice that the curves in Fig.~\ref{threshold_range} and Fig.~\ref{threshold_density} are mostly flat except for very high values of $T^{(c)}$. This implies that there is no need for additional resources to achieve a higher network-wide information dissemination, which reaffirms the fact that the intra-layer information dissemination is a much stricter condition than the network-wide information dissemination. Similarly, in Fig.~\ref{Fig_threshold_individual}, we fix the network-wide information dissemination threshold to $T^{(c)} = 0.5$ and observe the change in optimal parameters with a change in intra-layer information dissemination threshold. In this case, the device densities are increased successively followed by the communication ranges as the threshold for the proportion of simultaneously informed devices is increased.





\section{Conclusion}
In this paper, we have presented a generic framework for secure and reconfigurable design of IoT empowered battlefield networks. The framework provides a tractable approach to tune the physical network parameters to achieve the desired real-time data dissemination among different types of battlefield devices according to the assigned missions. It takes into account the perceived threat level from the opponent as well as the costs involved in the deployment and operation of combat equipment to provide a robust and cost effective design of communication networks in battlefields which can be highly useful in military planning. Optimized network parameters are provided for the two typical mission scenarios of intelligence and encounter battle in which the desired information spreading objectives are different from one another. Results show that the mission goals can be achieved by either appropriately changing the deployment density of combat units or by changing their transmission powers or both in response to changing mission requirements. Moreover, the network can be reconfigured according to the periodic assessment of lost connectivity or changes in security requirements to meet the desired mission goals.


\appendices
\section{Proof of Lemma~\ref{lemma_intra_layer_degree}} \label{proof_lemma_intra_layer_degree}
To evaluate the degree distribution of the devices in the first network layer, we proceed as follows:
{
\begin{align}\label{K_1_joint}
\mathbb{P}(K_1 = k) = \sum_{j = 1}^2 \mathbb{P}(K_1 = k | \mathcal{T}_i = j)  \mathbb{P}(\mathcal{T}_i = j),
\end{align}}
Since type-$\RN{2}$ devices are not active in the first network layer, so we know that $\mathbb{P}(K_1 = 0 | \mathcal{T}_i = 2) = 1$ and $\mathbb{P}(K_1 > 0 | \mathcal{T}_i = 2) = 0$. However, type-$\RN{1}$ devices can communicate with other type-$\RN{1}$ devices in the first network layer. Since the active devices in the first layer are also represented as a PPP, the degree of the RGG formed among the active devices is Poisson distributed~\cite{haenggi_sg}, i.e., $\mathbb{P}(K_1 = k | \mathcal{T}_i = 1) \sim \text{Poisson}(\lambda_1 \pi r_1^2)$. It follows that $\mathbb{P}(K_1 = 0 | \mathcal{T}_i = 1) = e^{- \lambda_1 \pi r_1^2}$ and $\mathbb{P}(K_1 > 0 | \mathcal{T}_i = 1) = \frac{e^{- \lambda_1 \pi r_1^2}(\lambda_1 \pi r_1^2)^k}{k!}, k >0$. Finally, since $\mathbb{P}(\mathcal{T}_i = 1) = p$ and $\mathbb{P}(\mathcal{T}_i = 2) = 1 - p$, we can substitute the corresponding expressions for $K_1 = 0$ and $K_1 > 0$ in~\eqref{K_1_joint} to obtain the result in~\eqref{K_1_pdf}. Similarly, for the distribution of $K_2$, we know that $\mathbb{P}(K_2 = k) \sim \text{Poisson}(\lambda_2 \pi r_2^2)$ regardless of $\mathcal{T}_i$. This directly leads to the result in~\eqref{K_2_pdf}. Using these probability distributions, it is straightforward to derive the expected degree in both layers given by~\eqref{E_K1} and~\eqref{E_K2}.

\section{Proof of Lemma~\ref{combined_degree_lemma}} \label{combined_degree_lemma_proof}
\begin{figure}
  \centering
  \includegraphics[width=2in]{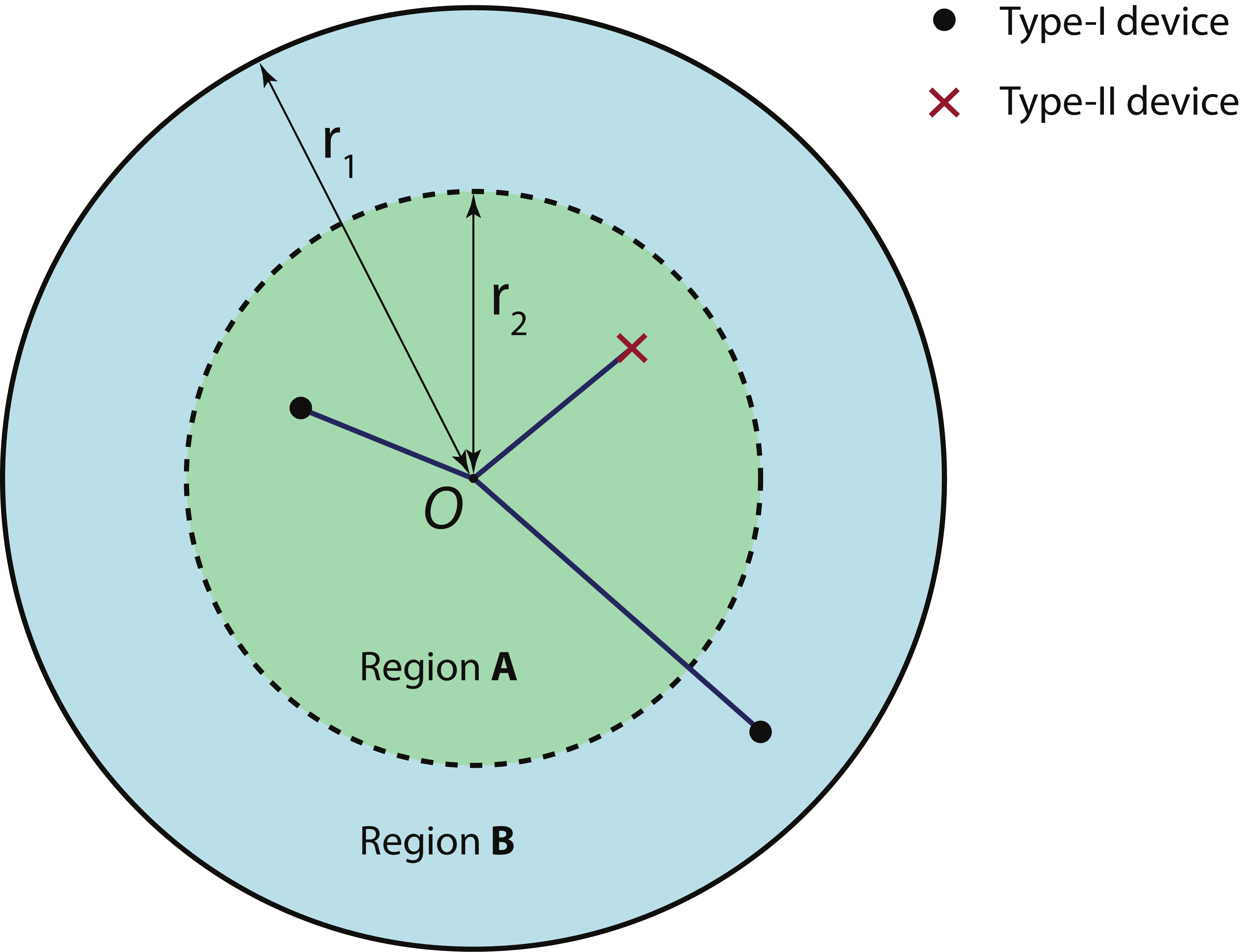}\\
  \caption{Combined degree of a typical device located at the origin.}\label{combined_degree_diagram}
\end{figure}

\noindent The distribution of the combined degree can be evaluated as follows:
{
\begin{align}\label{K_c_proof}
\mathbb{P}(K_c = k) = \sum_{j = 1}^2 \mathbb{P}(K_c = k | \mathcal{T}_i = j)  \mathbb{P}(\mathcal{T}_i = j),
\end{align}}
\textcolor{black}{It is clear that $\mathbb{P}(K_c = k | \mathcal{T}_i = 2) \sim \text{Poisson}(\lambda_2 \pi r_2^2)$ as type-$\RN{2}$ devices are only connected in the second network layer. However, if a typical device is of type-$\RN{1}$, then the degree needs to be more carefully evaluated. We have provided an illustration in Fig.~\ref{combined_degree_diagram} to aid in characterizing the degree. The degree of a type-$\RN{1}$ device placed at the origin can be expressed as $K_c = 2N_{1\mathbf{A}} + N_{2\mathbf{A}} + N_{1\mathbf{B}}$, where $N_{1\mathbf{A}}$ and $N_{2\mathbf{A}}$ represent the number of devices of type-$\RN{1}$ and type-$\RN{2}$ respectively in the green shaded circular region $\mathbf{A}$. $N_{1\mathbf{B}}$ represents the number of type-$\RN{1}$ devices in the blue shaded hollow circular region $\mathbf{B}$. Note that type-$\RN{1}$ and type-$\RN{2}$ devices are distributed according to independent PPPs with intensity $p \lambda$ and $(1-p)\lambda$. It results in the fact that $N_{1\mathbf{A}}$, $N_{2\mathbf{A}}$, and $N_{1\mathbf{B}}$ are independent random variables following a $\text{Poisson}(p \lambda \pi r_2^2)$, $\text{Poisson}((1-p) \lambda \pi r_2^2)$, and $\text{Poisson}(p \lambda \pi (r_1^2 - r_2^2))$ respectively. The sum $K_c$ is not composed of independent terms, however, due to $N_{1\mathbf{A}}$ which has a multiplicity of $2$. Therefore, $\mathbb{P}(K_c = k | \mathcal{T}_i = 1) = \mathbb{P}(N_{1\mathbf{A}} = \frac{k}{2}) * \mathbb{P}(N_{2\mathbf{A}} = k ) * \mathbb{P}(N_{1\mathbf{B}} = k)$, where $*$ represents the convolution operator. Therefore, the probability distribution of the combined degree can be obtained by substituting the above developed expressions into~\eqref{K_c_proof} and using the fact that $\mathbb{P}(\mathcal{T}_i = 1) = p$ and $\mathbb{P}(\mathcal{T}_i = 2) = 1 - p$. The average combined layer degree can be obtained as $\mathbb{E}(K_c) = p \left( 2 \mathbb{E}(N_{1\mathbf{A}}) + \mathbb{E}(N_{2\mathbf{A}}) + \mathbb{E}(N_{1\mathbf{B}}) \right) + (1 - p)(\lambda \pi r_2^2) = p \lambda_1 \pi r_1^2 + \lambda_2 \pi r_2^2$.
}


\section{Proof of Theorem~\ref{ex_and_uniq}} \label{existence}
To prove that the fixed point equation described by \eqref{fixed-point-equation} has a unique solution in the domain $\Theta^{(c)} > 0$, we make use of the Banach fixed-point theorem (or contraction mapping theorem)~\cite{contraction_mapping}. We prove that the functional
{
\begin{align}
F(\Theta(\alpha^{(c)})) = \frac{1}{\mathbb{E}[K_c]}\mathbb{E}\left[\frac{K_c^2 \alpha^{(c)} \Theta(\alpha^{(c)})}{1 + K_c \alpha^{(c)} \Theta(\alpha^{(c)})}\right]
\end{align}}
experiences a contraction for all $\Theta(\alpha^{(c)}) \in [0,1]$. More precisely, we prove that $|F(\Theta_1(\alpha^{(c)})) - F(\Theta_2(\alpha^{(c)}))| \leq c|\Theta_1(\alpha^{(c)}) - \Theta_2(\alpha^{(c)})|$ for any $\Theta_1(\alpha^{(c)}),\Theta_2(\alpha^{(c)}) \in [0,1]$, where $0 \leq c < 1$. The fact that the constant $c$ is strictly less than $1$ implies that the functional is contracted. The proof is as follows:
{
\begin{align}
&\left|F(\Theta_1(\alpha^{(c)})) - F(\Theta_2(\alpha^{(c)}))\right| = \\ & \left|\frac{1}{\mathbb{E}[K_c]}\mathbb{E}\left[\frac{K_c^2 \alpha^{(c)} \Theta_1(\alpha^{(c)})}{1 + K_c \alpha^{(c)} \Theta_1(\alpha^{(c)})}\right] - \right. \notag  \\ & \hspace{1.2in}\left. \frac{1}{\mathbb{E}[K_c]}\mathbb{E}\left[\frac{K_c^2 \alpha^{(c)} \Theta_2(\alpha^{(c)})}{1 + K_c \alpha^{(c)} \Theta_2(\alpha^{(c)})}\right]\right|, \notag \\
&= \frac{\left|  \Theta_1(\alpha^{(c)}) - \Theta_2(\alpha^{(c)})  \right|}{\mathbb{E}[K_c]} \times \notag \\ & \mathbb{E} \left[ \frac{K_c^2 \alpha^{(c)}}{(1 + K_c \alpha^{(c)} \Theta_1(\alpha^{(c)}))(1 + K_c \alpha^{(c)} \Theta_2(\alpha^{(c)}))} \right].
\end{align}}
To complete the proof, we need to show that
{
\begin{align}
\frac{1}{\mathbb{E}[K_c]} \mathbb{E} \left[ \frac{K_c^2 \alpha^{(c)}}{(1 + K_c \alpha^{(c)} \Theta_1(\alpha^{(c)}))(1 + K_c \alpha^{(c)} \Theta_2(\alpha^{(c)}))} \right] < 1,
\end{align}}
Let $g(K_c) = \frac{K_c^2 \alpha^{(c)}}{(1 + K_c \alpha^{(c)} \Theta_1(\alpha^{(c)}))(1 + K_c \alpha^{(c)} \Theta_2(\alpha^{(c)}))}$. It can be proved that $g(K_c)$ is concave for $K_c \geq 0$ by showing that $g^{\prime \prime}(K_c) < 0, \forall K_c \geq 0$. Therefore, using Jensen's inequality~\cite{jensen}, we can say that $\mathbb{E}[g(K_c)] \leq g(\mathbb{E}[K_c])$, with equality iff $K_c$ is deterministic. It follows that
{
\begin{align}
&\frac{1}{\mathbb{E}[K_c]} \mathbb{E} \left[ \frac{K_c^2 \alpha^{(c)}}{(1 + K_c \alpha^{(c)} \Theta_1(\alpha^{(c)}))(1 + K_c \alpha^{(c)} \Theta_2(\alpha^{(c)}))} \right] \leq \notag \\
 &\frac{\mathbb{E}[K_c] \alpha^{(c)}}{(1 + \mathbb{E}[K_c] \alpha^{(c)} \Theta_1(\alpha^{(c)}))(1 + \mathbb{E}[K_c] \alpha^{(c)} \Theta_2(\alpha^{(c)}))}, \notag \\
 &= \frac{1}{\Theta_1(\alpha^{(c)}) + \Theta_2(\alpha^{(c)}) + \mathbb{E}[K_c]\alpha^{(c)}\Theta_1(\alpha^{(c)})\Theta_2(\alpha^{(c)}) + \frac{1}{\mathbb{E}[K_c]\alpha^{(c)}}    }. \label{condition_contraction}
\end{align}}
\noindent The expression in~\eqref{condition_contraction} is strictly less than $1$ only if the following condition is satisfied:
{
\begin{align}\hspace{-0.1in}
\Theta_1(\alpha^{(c)}) + \Theta_2(\alpha^{(c)}) + \mathbb{E}[K_c]\alpha^{(c)}\Theta_1(\alpha^{(c)})\Theta_2(\alpha^{(c)}) + \notag \\ \frac{1}{\mathbb{E}[K_c]\alpha^{(c)}} > 1. \label{condition_contraction2}
\end{align}}
The condition in~\eqref{condition_contraction2} depends on the relative magnitudes of the quantities $\mathbb{E}[K_c]$ and $\alpha^{(c)}$. Regardless, it reveals that we need to exclude the values of $\Theta(\alpha^{(c)})$ that are too close to zero. For sufficiently large values of $\Theta(\alpha^{(c)})$, it is clear from \eqref{condition_contraction2}, that $F(\Theta(\alpha^{(c)}))$ is indeed a contraction with respect to the absolute value metric. Hence, by the contraction mapping theorem, $F(\Theta(\alpha^{(c)}))$ has a unique fixed point in the domain $\Theta(\alpha^{(c)}) > 0$.

The nonzero equilibrium solution can be obtained by solving the following equation:
{
\begin{align}
1 = \frac{1}{\mathbb{E}[K_c]} \mathbb{E} \left[ \frac{K_c^2 \alpha^{(c)}}{1 + K_c \alpha^{(c)} \Theta(\alpha^{(c)})} \right]. \label{non_zero_eq}
\end{align}}
Let $h(\Theta(\alpha^{(c)})) = \frac{1}{\mathbb{E}[K_c]} \mathbb{E} \left[ \frac{K_c^2 \alpha^{(c)}}{1 + K_c \alpha^{(c)} \Theta(\alpha^{(c)})} \right]$. We need to find a solution to the equation $h(\Theta(\alpha^{(c)})) = 1$ in the domain $0 < \Theta(\alpha^{(c)}) \leq 1$. It is clear that $h(\Theta(\alpha^{(c)}))$ is monotonically decreasing for $\Theta(\alpha^{(c)}) > 0$. Therefore, it is sufficient to show that $h(0) > 1$ and $h(1) < 1$ for a unique nonzero solution to exist for the equation $h(\Theta(\alpha^{(c)})) = 1$. This result is proved below:
{
\begin{align}
h(0) &= \frac{1}{\mathbb{E}[K_c]} \mathbb{E} \left[ K_c^2 \alpha^{(c)} \right] = \alpha^{(c)} \frac{\mathbb{E}[K_c^2]}{\mathbb{E}[K_c]}, \\
h(1) &= \frac{1}{\mathbb{E}[K_c]} \mathbb{E} \left[ \frac{K_c^2 \alpha^{(c)}}{1 + K_c \alpha^{(c)}}  \right] \notag \\ & = \frac{1}{\mathbb{E}[K_c]} \mathbb{E} \left[ K_i\frac{K_c \alpha^{(c)}}{1 + K_c \alpha^{(c)}} \right]  < \frac{1}{\mathbb{E}[K_c]} \mathbb{E}[K_c] = 1.\label{h_1}
\end{align}}
In~\eqref{h_1}, the inequality follows from the fact that $\frac{K_c \alpha^{(c)}}{1 + K_c \alpha^{(c)}} < 1, \forall K_c > 0, \alpha^{(c)} > 0$. A nonzero solution to \eqref{non_zero_eq} exists only if $h(0) \geq 1$, which implies that $\alpha^{(c)} \geq \frac{\mathbb{E}[K_c]}{\mathbb{E}[K_c^2]}$. This is exactly the critical spreading rate, also known as epidemic threshold, in the SIS model~\cite{epidemic_scale_free}.

\section{Proof of Theorem~\ref{theorem_solution}} \label{solution}
\begin{figure}
  \centering
  \includegraphics[width=2.5in]{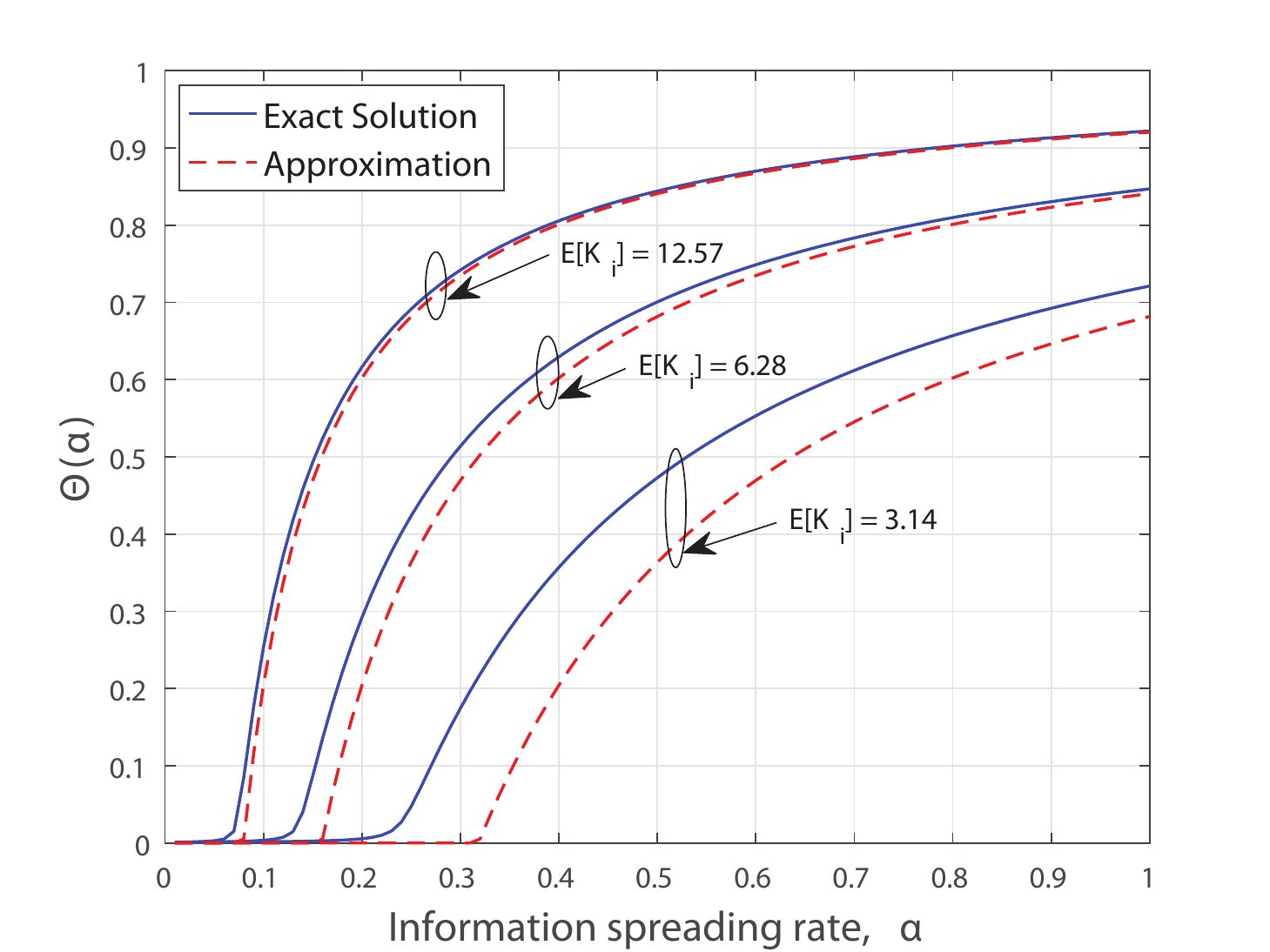}\\
  \caption{Accuracy of Jensen's lower bound. }\label{approx_accuracy}
\end{figure}

Obtaining the nonzero solution for the fixed point equation \eqref{fixed-point-equation} in closed form is not possible since we need to solve the following equation for $\Theta(\alpha^{(c)})$:
{
\begin{align}
1 = \frac{1}{\mathbb{E}[K_c]} \sum_{k=0}^{\infty} \frac{K_c^2 \alpha^{(c)}}{1 + K_c \alpha^{(c)} \Theta(\alpha^{(c)})} \mathbb{P}(K_c = k), \label{fixed_point_exact}
\end{align}}
\textcolor{black}{where $P(K_c = k)$ is difficult to obtain in closed form}.
Therefore, we resort to finding an approximation for the solution which is asymptotically accurate. Let $g(K_c) = \frac{K_c^2 \alpha^{(c)}}{1 + K_c \alpha^{(c)} \Theta(\alpha^{(c)})}$. Since $g^{\prime \prime}(K_c) > 0, \forall K_i \geq 0$, so $g(K_i)$ is a convex function for $K_i \geq 0$. Using Jensen's inequality, we can say that $\mathbb{E}[g(K_c)] \geq g(\mathbb{E}[K_c])$, with equality only if $K_c$ is deterministic. This implies the following:
{
\begin{align}
\mathbb{E} \left[ \frac{K_c^2 \alpha^{(c)}}{1 + K_c \alpha^{(c)} \Theta(\alpha^{(c)})} \right] \geq \frac{\mathbb{E}[K_c]^2 \alpha^{(c)}}{1 + \mathbb{E}[K_c] \alpha^{(c)} \Theta(\alpha^{(c)})}.
\end{align}}
Therefore, we can write~\eqref{fixed_point_exact} as follows:
{
\begin{equation}
1 \geq \frac{\mathbb{E}[K_c] \alpha^{(c)}}{1 + \mathbb{E}[K_c] \alpha^{(c)} \Theta(\alpha^{(c)})},
\end{equation}}
which leads to the final solution,
{
\begin{align}
\Theta(\alpha^{(c)}) \geq 1 - \frac{1}{\alpha^{(c)} \mathbb{E}[K_c]}.
\end{align}}
Using our prior knowledge that $\Theta(\alpha^{(c)}) \geq 0$, we need to ensure that $\alpha^{(c)} \mathbb{E}[K_c] \geq 1$. In general, the complete solution can be expressed as $\Theta(\alpha^{(c)}) \geq \max(0, 1 - \frac{1}{\alpha^{(c)} \mathbb{E}[K_c]})$.
To measure the accuracy of this bound, we solve the fixed-point equation exactly using an fixed-point iteration and compare the results for different values of $\alpha^{(c)}$ and $\mathbb{E}[K_c] = \lambda_i \pi r_i^2$. We choose a fixed $r_i = 0.2$ km and $\lambda_i = [25, 50, 100]$ km$^{-2}$, which results in $\mathbb{E}[K_i] = [3.14, 6.28, 12.57]$. A plot of the results is provided in Fig.~\ref{approx_accuracy}. It can be observed that the lower bound obtained using Jensen's inequality is tight for all values of $\alpha^{(c)}$ when $\mathbb{E}[K_i] \gg 1$. 

\section{Proof of Theorem~\ref{thm2}} \label{thm2_proof}
The objective is to obtain a fixed point solution to the equations \eqref{fp1} and \eqref{fp2}. Since the equations are decoupled, we can employ the same approach used for Theorem~\ref{theorem_solution} to show the existence and uniqueness of the fixed point. Consequently, a lower bound approximation of the solution for each fixed-point equation can be obtained using the Jensen's inequality as in \textbf{Appendix~\ref{solution}}. The condition for existence of a solution is $\alpha^{(c)} \geq \frac{\mathbb{E}[K_1]}{\mathbb{E}[K_1^2]}$ and $\alpha^{(c)} \geq \frac{\mathbb{E}[K_2]}{\mathbb{E}[K_2^2]}$, which can be written as $\alpha^{(c)} \geq \max \left( \frac{\mathbb{E}[K_1]}{\mathbb{E}[K_1^2]}, \frac{\mathbb{E}[K_2]}{\mathbb{E}[K_2^2]} \right)$.
%

\bibliographystyle{IEEEtran}
\bibliography{references}

\begin{IEEEbiography}
    [{\includegraphics[width=1in,height=1.25in,clip,keepaspectratio]{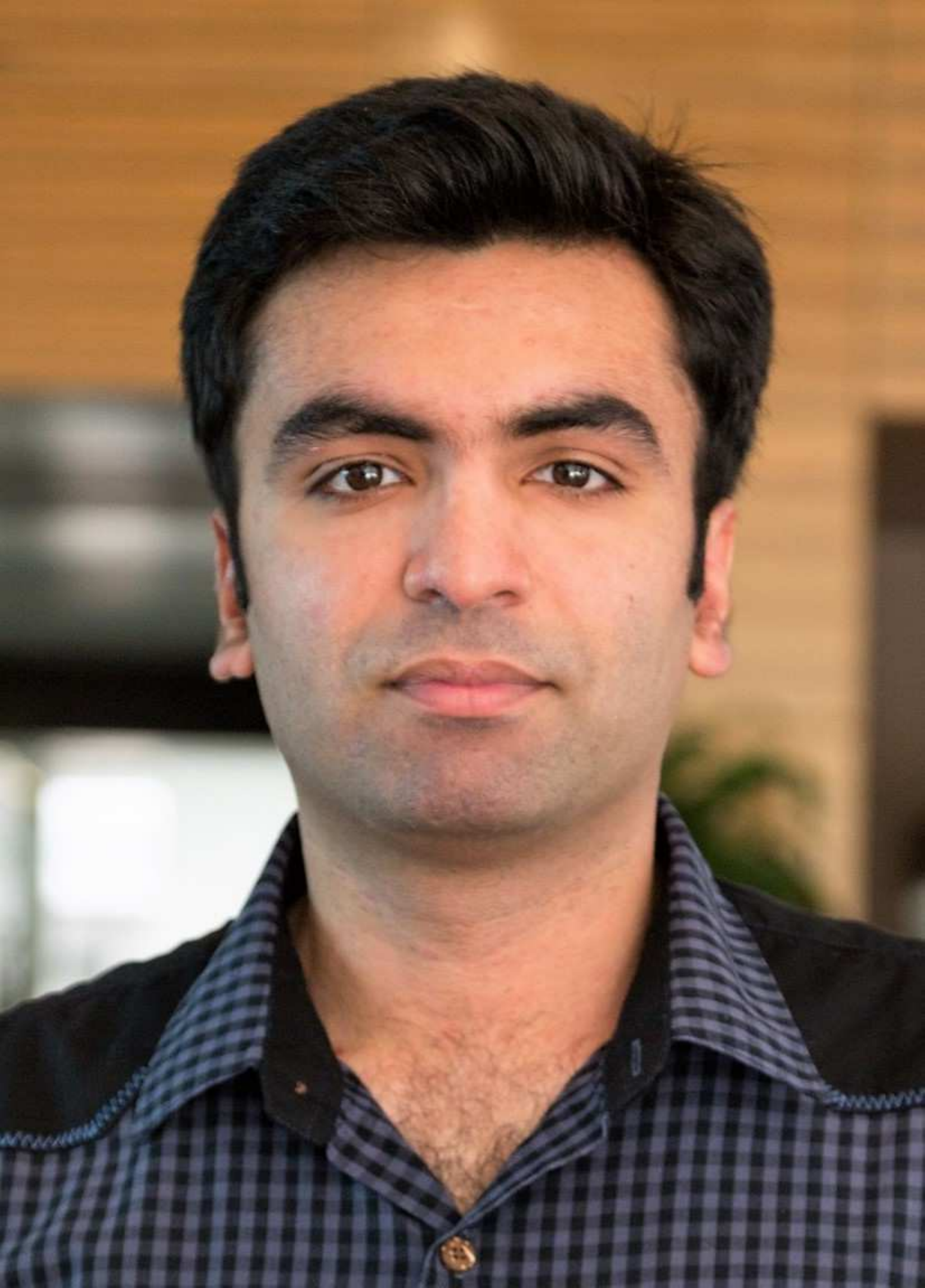}}]{Muhammed Junaid Farooq} received the B.S. degree in electrical engineering from the School of Electrical Engineering and Computer Science (SEECS), National University of Sciences and Technology (NUST), Islamabad, Pakistan, the M.S. degree in electrical engineering from the King Abdullah University of Science and Technology (KAUST), Thuwal, Saudi Arabia, in 2013 and 2015, respectively. Then, he was a Research Assistant with the Qatar Mobility Innovations Center (QMIC), Qatar Science and Technology Park (QSTP), Doha, Qatar. Currently, he is a PhD student at the Tandon School of Engineering, New York University (NYU), Brooklyn, New York. His research interests include modeling, analysis and optimization of wireless communication systems, cyber-physical systems, and the Internet of things. He was the recipient of the President's Gold Medal for the best academic performance from the National University of Sciences and Technology (NUST).
\end{IEEEbiography}

\begin{IEEEbiography}
    [{\includegraphics[width=1in,height=1.25in,clip,keepaspectratio]{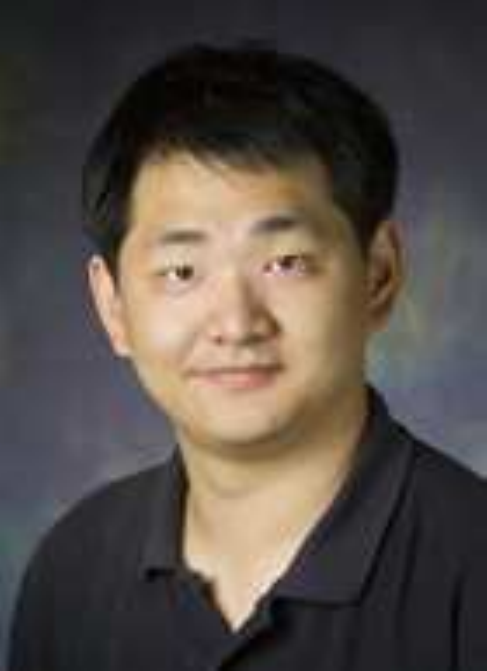}}]{Quanyan Zhu} (S'04, M'12) received B. Eng. in Honors Electrical Engineering from McGill University in 2006, M.A.Sc. from University of Toronto in 2008, and Ph.D. from the University of Illinois at Urbana-Champaign (UIUC) in 2013. After stints at Princeton University, he is currently an assistant professor at the Department of Electrical and Computer Engineering, New York University. He is a recipient of many awards including NSERC Canada Graduate Scholarship (CGS), Mavis Future Faculty Fellowships, and NSERC Postdoctoral Fellowship (PDF). He spearheaded and chaired INFOCOM Workshop on Communications and Control on Smart Energy Systems (CCSES), and Midwest Workshop on Control and Game Theory (WCGT). His current research interests include Internet of things, cyber-physical systems, security and privacy, and system and control.
\end{IEEEbiography}

\end{document}